\documentclass[11pt]{article}
\usepackage{jheppub}
\usepackage[T1]{fontenc}

\usepackage{amssymb}
\usepackage{amsfonts}
\usepackage{amsbsy}
\usepackage{amsmath}
\usepackage{amsthm}
\usepackage{graphicx}
\usepackage{epstopdf}
\usepackage[vcentermath]{youngtab}
\usepackage{multirow}
\usepackage{latexsym} 
\usepackage{array}
\usepackage{color}
\definecolor{Red}{rgb}{1.00, 0.00, 0.00}
\definecolor{Blue}{rgb}{0.00, 0.00, 1.00}
\definecolor{Purple}{cmyk}{0.45,0.86,0,0}%%%PANTONE PURPLE

\newtheorem{theorem}{Theorem}[section]

\def\Z{\mathbb{Z}}
\def\F{\mathbb{F}}
\def\Q{\mathbb{Q}}

\def\P{\mathbb{P}}

\def\n3a{t}

\def\tr{{\mathrm{tr}}}

\def\phit{\phi_0}
\def\phif{\phi_1}

\def\torus{\boldsymbol{T}}

\def\age{{\mathfrak{e}}}

\def\agf{{\mathfrak{f}}}

\newcommand{\gsu}[0]{{\mathfrak{su}}}
\newcommand{\gso}[0]{\mathfrak{so}}
\newcommand{\gsp}[0]{\mathfrak{sp}}
\newcommand{\gggg}[0]{\mathfrak{g}}

\title{\boldmath
Charge completeness\\ and  the massless charge lattice\\
in F-theory models of supergravity
}

\author[1]{David R. Morrison}
\author[]{and}
\author[2]{Washington Taylor}

\affiliation[1]{Departments of Mathematics and Physics\\
University of California, Santa Barbara\\
Santa Barbara, CA 93106, USA}
\affiliation[2]{Center for Theoretical Physics\\
Department of Physics\\
Massachusetts Institute of Technology\\
77 Massachusetts Avenue\\
Cambridge, MA 02139, USA}

\emailAdd{drm at math.ucsb.edu}
\emailAdd{wati at mit.edu}

\preprint{\today \\
\hfill
MIT-CTP-5172, UCSB-MATH-2021-03}

\abstract{
We prove that, for every 6D supergravity theory that has an F-theory
description, the property of charge completeness for the connected component
of the gauge group (meaning that all charges in the corresponding
charge lattice are realized by massive or massless states in the
theory) is equivalent to a standard assumption made in F-theory for
how geometry encodes the global gauge theory by means of the
Mordell-Weil group of the elliptic fibration.  
This result also holds in 4D F-theory constructions for the
parts of the gauge group that come from  sections and from
7-branes.  We find that in
many 6D F-theory models the full charge lattice of the theory is
generated by massless charged states; this occurs for each gauge
factor where the associated anomaly coefficient satisfies a simple
positivity condition.  We describe many of the cases where this
massless charge sufficiency condition holds, as well as exceptions
where the positivity condition fails, and analyze the related global
structure of the gauge group and associated Mordell-Weil torsion in
explicit F-theory models.
 }
  
\begin{document}
\maketitle

\flushbottom

%--------------------------------
\section{Introduction}

String theory, in its many avatars, gives rise to a wide range of
vacuum solutions in various dimensions with a great variety of
different gauge groups and matter content coupled to quantum gravity.
This large range of solutions has led to the suggestion that
essentially ``anything goes,'' and that virtually any consistent
quantum field theory can be found coupled to quantum gravity in some
string theory.  In fact, however, quantum gravity and string theory
place strong constraints on the structure of theories with a
consistent UV completion.

This tension between the constraints coming from string theory and
consistency conditions apparent for a low-energy quantum field theory
coupled to gravity has been present since the early days of string
theory.  A simple example is the set of consistent ${\cal N} = 1$
supersymmetric gravity theories in ten dimensions.  The seminal work
of Green and Schwarz \cite{Green-Schwarz} showed that all such
supergravity theories were inconsistent except those with gauge groups
Spin$(32)/\Z_2$, $E_8 \times E_8$, $E_8 \times U(1)^{248}$ and
$U(1)^{496}$.  At the time only the first of these gauge groups was
known to arise in string theory.  Motivated by this result, the
``string quartet'' identified the heterotic $E_8 \times E_8$ string
theory \cite{string-quartet}.  More recently, Vafa coined the term
``swampland'' \cite{Vafa-swampland}
to describe those theories that appear to be consistent
under all known quantum consistency conditions but that are not
(known to be)
realized in string theory, and pointed out that,  at the time this term was coined,  the 10D
theories with gauge groups $E_8 \times U(1)^{248}$ and $U(1)^{496}$
appeared to be in the swampland.  In \cite{10D-universality}, it was
shown that in fact these theories cannot be simultaneously compatible
with supersymmetry and gauge invariance.  The same conclusion was
derived from a string worldsheet point of view in \cite{Kim-Shiu-Vafa}.  Thus,
for minimally supersymmetric 10D theories of gravity coupled to gauge groups, 
at least at the level of massless spectra the swampland
is empty.  In fewer than ten space-time dimensions, however, the set of
theories with known realizations in string theory is a proper subset of those that
satisfy all known consistency conditions, and the ongoing study of the
difference between these two sets has been a productive focus of much
recent research effort; an overview of recent conjectures and
developments related to the swampland is given in \cite{Palti:2019pca}.

Six-dimensional ${\cal N} = (1, 0)$ supergravity theories present a
particularly interesting domain in which to investigate ``swampland''
related questions.  As in ten dimensions, the requirement of
gravitational and mixed gauge-gravitational anomaly cancellation
provides strong constraints on the set of possibly consistent 6D
theories.  At the same time, string constructions of 6D supergravity
theories are fairly well understood, particularly in the context of
F-theory \cite{Vafa-F-theory, Morrison-Vafa-I, Morrison-Vafa-II}.  The
set of 6D supergravity theories that can be realized through known
string constructions 
essentially forms one large moduli space, with
different branches connected by various kinds of geometric
transitions.\footnote{Note, it has not been shown how the theories in
  \cite{frozen} are connected to the connected moduli space of
  standard F-theory constructions.}  It was conjectured in
\cite{6D-universality} that, as in 10D, it may be possible to show
that every quantum-consistent massless 6D supergravity spectrum is
realized in string theory.  The close connection between the structure
of 6D supergravity and F-theory (see e.g.\ \cite{KMT-II}) gives many
insights into these theories, but at this time there are still
substantial classes of apparently consistent 6D supergravity theories
without an F-theory realization; for example, there is an infinite
class of theories with no tensor multiplets, a single U(1) gauge
factor, and matter fields with arbitrarily large U(1) charges that
cannot be realized in F-theory but that satisfy anomaly cancellation
and all other known quantum consistency conditions on low-energy 6D
field theories coupled to gravity \cite{TT-infinite}. Other aspects of
the 6D supergravity swampland have been explored in a variety of
papers, including \cite{Seiberg-Taylor, frozen, Monnier-Moore,
  Kim-Shiu-Vafa, Lee-Weigand, Angelantonj:2020pyr, Raghuram:2020vxm, Tarazi:2021duw}.

Another aspect of ``swampland'' related questions involves certain
features that appear to be typical of all string constructions (in a
variety of dimensions), and for some of which there are heuristic
arguments involving quantum gravity features such as black hole
information and radiation.  Some analysis and
conjectures based on features of this
type were given in \cite{Adams:2006sv, Ooguri-Vafa, ArkaniHamed:2006Dz}.  One particular
feature of string theory is the general aspect that any time there is
a symmetry, with an associated gauge or higher form field, charged
objects appear in the theory, and these objects saturate the allowed
space of charges in most (perhaps all) known string solutions. This
principle was part of the philosophy underlying the development of
D-branes \cite{Polchinski-completeness}, which are the charged objects
associated with the Ramond-Ramond fields of type II string theory.
This notion of ``charge completeness'' and the related idea that
theories of quantum gravity cannot have global symmetries, only gauge
symmetries, were articulated clearly in \cite{Banks-Seiberg}.  For theories in anti-de Sitter space with holographic dual
descriptions, these conjectures were  recently proven by Harlow and Ooguri
\cite{Harlow-Ooguri}.

In this paper we address the charge completeness question in the
context of 6D and 4D supergravity theories.  We prove using
Poincar\'{e} duality that for 6D ${\cal N} = (1, 0)$ supergravity
theories that arise from F-theory, the charge completeness
hypothesis for the connected component of the gauge group
 is equivalent to the standard assumption made in F-theory for how the
 global structure of the gauge group is encoded in geometry.\footnote{
As described in more detail in Section \ref{sec:F-theory} and the Appendix,
the
``standard assumption'' 
\cite{Aspinwall-Morrison,LieF,torsion-MW} 
relates the fundamental group $\pi_1(G)$ of
the gauge group $G$ to the group of sections (the ``Mordell--Weil group'')
of the elliptic fibration used in the F-theory construction, by considering
which representations of the corresponding Lie algebra can occur
in the presence of Mordell--Weil group elements.}
We  show that
the same result also holds in 4D F-theory constructions for the part of the
gauge group that is realized through 7-branes and sections.  (Note that in 6D all
of these supergravity theories live in flat space-time, so that the
proof of Harlow and Ooguri is not applicable).

We furthermore observe that for many 6D ${\cal N} = (1, 0)$
supergravity theories a stronger condition holds: in these models the charge
lattice is completely determined by the set of massless fields in the
theory.  We identify a simple positivity condition on the anomaly
coefficients of the gauge factors that seems to be sufficient to imply
this ``massless charge sufficiency'' condition.  We give a large set
of examples of 6D supergravity theories with connected nonabelian and
abelian gauge groups where this condition holds, and analyze some of
the cases where it does not. Mordell-Weil torsion and the global
structure of the gauge group play an important role in understanding
these structures.

The structure of the paper is as follows.  We begin in Section
\ref{sec:6D} with a very brief review of 6D ${\cal N} = (1, 0)$
supergravity theories and their realization in F-theory.  In Section
\ref{sec:hypotheses} we state the completeness and massless charge
sufficiency conditions.  We prove a version of the completeness
hypothesis for those theories with an F-theory realization, and
conjecture that the massless charge sufficiency condition holds in all
but certain special exceptional circumstances.  In Section
\ref{sec:nonabelian}, we give a variety of examples of 6D ${\cal N} =
(1, 0)$ supergravity theories with nonabelian gauge groups where the
gauge algebra does not uniquely determine the global structure of the
gauge group.  In many of these theories we prove 
that the massless
charges in the theory are sufficient to generate the full charge
lattice, and explicitly describe the connection in these theories
between Mordell-Weil torsion in the F-theory model and the fundamental
group of the gauge group of the associated 6D supergravity theory.  In
Section \ref{sec:abelian} we consider an example with an abelian gauge
group, and in Section \ref{sec:special-cases} we describe several of
the exceptional classes of models where the massless charge spectrum
is not sufficient to determine the full charge lattice of the theory.
Section \ref{sec:conclusions} contains some concluding remarks.
We include an appendix with a discussion of the topology of the gauge
group and its relation to representation theory.

Some recent papers touch on issues closely related to the subject and
techniques of this work.  The structure of discrete 1-form symmetries
in 6D supergravity theories with Mordell-Weil torsion like many of the
models studied here was investigated in \cite{Apruzzi:2020zot}.  In
\cite{Cvetic:2020kuw} related aspects of the global structure of the
gauge group for 8D F-theory models have been investigated.  In
\cite{Raghuram:2020vxm}, a general characterization is given of a
large set of 6D supergravity models that appear to be compatible with
known quantum consistency conditions but lack an F-theory realization.
Such models violate an automatic enhancement condition that seems to
hold for all theories that come from F-theory.  The automatic
enhancement conjecture made in that paper is closely related in some
cases to the massless charge sufficiency conjecture presented here, as
discussed further in that paper; the analysis of that paper was
inspired in part by some examples encountered in this work.

\section{Quick review of 6D supergravity and F-theory realizations}
\label{sec:6D}

We begin by briefly reviewing the structure of 6D ${\cal N} =  (1, 0)$
supergravity theories and their realizations in F-theory

\subsection{Anomaly conditions}

In general a 6D ${\cal N} =  (1, 0)$ supergravity theory has $T$
tensor multiplets, $V =\dim G = \dim G^0$ vector multiplets   where the connected
component\footnote{The gauge group $G$ can also have a disconnected part $G/G^0$ but that
will not concern us in this paper.  In fact, $\pi_1(G)=\pi_1(G^0)$ so the
condition as we have stated it is unchanged by passing from $G$ to $G^0$.} 
$G^0$ of the gauge group $G$ can generally have both nonabelian and abelian
factors, taking the form\footnote{See the appendix for more details
about how this form is arrived at.}  $G^0 = (G_0 \times U(1)^r)/\Xi$ (with
$G_0$ simply connected and
$\Xi$ a finite subgroup of the center of $G_0 \times U(1)^r$),
% WT: switched G_NA to G_0 to match later use
 and $H$ matter
hypermultiplets, which can be neutral or charged under the gauge group.
The 6D gravitational, nonabelian gauge, and gravitational-nonabelian gauge
anomaly cancellation conditions \cite{gs-west, Sagnotti} can be
written in the form \cite{KMT-II}
\begin{align}
H-V+29T & = 273\\
0 & =  B^{i}_{\rm adj}  - \sum_R x^i_{R} B^i_{R} \label{eq:B-condition}\\
a \cdot a & =  9 - T \\ 
-a \cdot b_i & =  \frac{1}{6} \lambda_i  \left( \sum_R
x^i_R A^i_R -  A^i_{\rm adj}\right)  \label{eq:ab-condition}\\
b_i\cdot b_i & = \frac{1}{3} \lambda_i^2 \left( \sum_R x_R^i C^i_R -
 C^i_{\rm adj}  \right) \\
b_i \cdot b_j & =  \lambda_i \lambda_j \sum_{RS} x_{RS}^{ij} A_R^i
A_S^j.
\label{eq:bij-condition}
\end{align}
Here, $a, b_i$ are the coefficients associated with the Green-Schwarz
terms $a \cdot B \wedge R \wedge R, b_i \cdot B \wedge F_i \wedge F_i$
in the action, with $F_i$ the $i$th nonabelian field strength; these
coefficients lie in a lattice $\Gamma$ of  signature (1, $T$).  
The quantities $x_R^i$ denote the number of hypermultiplets transforming in
representation $R$ under gauge group factor $G_i$, while the quantities $x^{ij}_{RS}$
denote the number of hypermultiplets transforming in representations $R, S$
under the factors $G_i, G_j$.
The
coefficients $A_R,B_R,C_R$ are defined through
\begin{align}
\tr_R\, F^2 & = A_R\,  \tr\, F^2 \\
\tr_R\, F^4 & = B_R\, \tr\, F^4+C_R (\tr\, F^2)^2 \,, \nonumber
\end{align}
where  $F$ is a field strength for the appropriate group factor
and $\tr$ is a normalized trace.
%and $x_R^i$ denotes the number of hypermultiplets transforming in
%representation $R$ under gauge group factor $G_i$, while $x^{ij}_{RS}$
%is the number of hypermultiplets transforming in representations $R, S$
%under the factors $G_i, G_j$.
Tables of the coefficients $A_R, B_R, C_R$ can be found in,
e.g., \cite{Erler,  kt-finite, KPT} (see also \cite[appendix C]{Grassi-Morrison-2}).
Note that for groups with no quartic Casimir (like SU(2), SU(3), and the exceptional
groups), $B_R = 0$ and the condition
(\ref{eq:B-condition}) is trivial.  As a simple example of the anomaly
conditions, for SU(2) we have $\lambda = 1$, and the anomaly
coefficients $A_R, C_R$ for the fundamental and adjoint
representations are $1, 1/2$ and $4, 8$ respectively.

For a theory with a single U(1) gauge factor, the associated
gauge-gravitational and pure gauge anomaly conditions are 
\begin{align}
-a \cdot \tilde{b} &= \frac{1}{6} \sum_{q > 0} x_q q^2\,,
    \label{eq:U1ACsqr} \\
\tilde{b} \cdot \tilde{b} &= \frac{1}{3} \sum_{q > 0} x_q q^4\,.
    \label{eq:U1ACquar}
\end{align}
Here, $\tilde{b}$ is the U(1) Green-Schwarz anomaly coefficient and
$x_q$ is the number of matter hypermultiplets of charge $q$ under the
U(1).
More details on anomaly conditions for theories with multiple U(1) and
mixed abelian-nonabelian gauge groups, 
can be found in \cite{Erler,
  Park-Taylor, Park}.
Some constraints on the anomaly coefficients $a, b, \tilde{b}$ are
determined in \cite{KMT-II, Monnier-Moore-Park, Monnier-Moore}.

\subsection{F-theory models of 6D and 4D supergravity theories}
\label{sec:F-theory}

A large class of 6D and 4D supergravity theories can be realized by
an F-theory compactification\footnote{For simplicity, we do not include
the F-theory compactifications with ``frozen'' singularities studied 
in \cite{frozen}, nor do we allow background fluxes.} using  
an elliptic Calabi-Yau threefold or fourfold $X$.  Such an
F-theory compactification is defined\footnote{In order to admit an effective
description as a gauge theory coupled to supergravity, the singularities in
the Weierstrass model cannot be too bad:  along any locus of complex codimension
two in the base, the multiplicities of $(f,g)$ must be less than $(4,6)$, or else
the F-theory model will include a superconformal sector
\cite{Seiberg-Witten,Morrison-Vafa-II,Heckman-Morrison-Vafa,DelZotto-Heckman-Tomasiello-Vafa}.}
 through a Weierstrass model for the 
elliptic fibration on  $X$,
\begin{equation}
 y^2 = x^3 + f x + g \,,
\label{eq:Weierstrass}
\end{equation}
where $f, g$ are sections of line bundles ${\cal O} (-4 K_B), {\cal O}
(-6 K_B)$ over a base $B$ that is a complex surface for a 6D theory or
a complex threefold for a 4D theory.  Such an F-theory
compactification can be thought of as a compactification of type IIB
string theory on the base $B$ with an axiodilaton field $\tau$ that
varies over $B$ and matches the ratio of periods $\tau$ of the elliptic
curve defined through (\ref{eq:Weierstrass}) at each point in $B$.  
More detailed introductions to F-theory models of 6D and 4D
supergravity can be found in \cite{Morrison-TASI, Denef-F-theory,
  WT-TASI, Weigand-TASI, whatF}; here we review a few of the
main points.
We focus principally on reviewing the structure of
6D F-theory models; 4D F-theory models are similar but have additional
complications, as we discuss in the last paragraph of this section.

The nonabelian gauge algebra of a 6D supergravity theory realized through
F-theory is
associated with Kodaira singularities \cite{Kodaira} in the elliptic
fibration over complex curves $C_i$ in $B$.  In the type IIB picture,
these are 7+1-dimensional objects (stacks of D7 branes or F-theory
generalizations thereof)  supporting nonabelian gauge
factors.  There is a close correspondence between the geometry of the
elliptic Calabi-Yau threefold over $B$ and the physics of the
resulting 6D theory.  In particular, there is a direct mapping between
 the anomaly coefficients $a, b$ of the 6D
theory and certain divisors in $B$.  The string charge lattice $\Gamma$ of the 6D theory
corresponds to the homology lattice $H_{1, 1}(B, \Z)$.  The anomaly
coefficient $a$ corresponds to the canonical class $K_B$, while $b_i$
corresponds to the class of the complex curve $C_i$.  Abelian anomaly coefficients
$\tilde{b}$ are described in a slightly more subtle but similar way in
terms of specific divisors in $X$.  The detailed relationships between
the geometry of elliptic Calabi-Yau threefolds and the structure of
the 6D theory, which relate geometric conditions to the 6D anomaly
constraints, are described in further detail in \cite{Sadov, Grassi-Morrison,
  KMT, KMT-II, Grassi-Morrison-2, Park, Grimm-Kapfer, Monnier-Moore}.
F-theory in 6D is often analyzed as a limit of M-theory (see
e.g.\ \cite{b-Grimm}), where in the M-theory picture the
compactification manifold for the Coulomb branch
is a smooth Calabi-Yau threefold $\tilde{X}$
associated with a resolution of the singularities of $X$.

While the Kodaira singularities in the elliptic fibration that arise
at codimension one in the F-theory base  $B$ directly encode the gauge
algebra of the corresponding 6D supergravity theory, the global
structure of the gauge group is somewhat more subtle.
Assume that a 6D supergravity theory has gauge group $G$ and comes
from an F-theory Weierstrass model that defines an elliptic Calabi-Yau
threefold $X$ with resolution $\tilde{X}$.  The Cartan generators of the
Lie algebra are associated with the (divisorial) 
components of the Kodaira fibers
of the codimension one singularities in the (resolved) fibration;
these are divisors in $H^{1, 1} (\tilde{X})$.  Other divisors are
associated with rational sections of the elliptic fibration;  because rational sections
can be ``added'' (using the group law of the elliptic fibration)
the rational sections
form a group\footnote{Because one chosen section $\sigma_\infty$, the ``zero-section,'' 
serves as the additive
identity for the group law on the elliptic fibers and for the Mordell-Weil group
itself, the map $\sigma \mapsto [\sigma]$ which sends a rational section to its
divisor class in $H^{1, 1} (\tilde{X})$ is not a group homorphism.  However, the
modified map
$\sigma \mapsto [\sigma-\sigma_\infty]$ {\em does}\/ define a group homomorphism.}
 called the Mordell-Weil group, which is a finitely generated abelian group
and therefore has the form $\Z^r
\times (MW)_{\text{tors}}$ where 
$r$ is the 
%number of U(1) factors in the 6D gauge group
rank of the group of sections 
and $(MW)_{\text{tors}}$ is the Mordell-Weil 
torsion.
The standard assumption in
F-theory is that the global structure of the 6D gauge group $G$ is
determined by the condition that the Mordell-Weil group coincides with 
fundamental group of the 
gauge group (see \cite{Aspinwall-Morrison} and footnote 4 in \cite{LieF}), 
although there is no proof of
this assumption.  One of the primary results of this paper is to show
that this assumption is equivalent to the charge completeness
hypothesis for 6D F-theory models.  In general, 
the connected part $G^0$ of the 6D
gauge group $G$ is  a compact Lie group, which can be written as a
quotient $G^0 = (G_0 \times U(1)^r)/\Xi$, where $G_0$ is simply
connected and represents the nonabelian part of $G$,
and $\Xi$ is a finite subgroup of the 
center of $G_0 \times U(1)^r$.  
The standard assumption regarding the
global structure of the gauge group is then that 
(1) $\pi_1(G)_{\text{tors}}$ coincides with 
$\Xi \cap G_0 = (MW)_{\text{tors}}$,  and (2)
the rank of the free part of the Mordell--Weil group 
measures the number of $U(1)$ factors in the gauge group.
Note that when $r = 0$, $G^0 = G_0/\Xi$, and $\Xi = (MW)_{\text{tors}}$; 
we consider a
number of cases of this type in Section \ref{sec:nonabelian}.
On the other hand, when there are abelian factors there can be a
nontrivial finite quotient even when $(MW)_{\text{tors}}$ is trivial;  
for example,
models having such structure with group $U(N)=(SU(N) \times U(1))/\Z_N$
or with the standard model gauge group
$(SU(3) \times SU(2) \times U(1))/\Z_6$ have been studied in
various places in the literature \cite{Lawrie-sw, Grimm-kk,
  Cvetic-Lin,
Lin-Weigand, Cvetic-kpor, Lin-Weigand-2,
  Cvetic-llo, Cvetic-hllt, TT-sm, RTT}.

The matter content of the 6D supergravity theory arising from an
F-theory model arises from a combination of local singularity
enhancements associated with intersecting 7-branes in the IIB picture
and nonlocal structure associated with  adjoint representations of
the gauge group \cite{Morrison-Vafa-II, Bershadsky-all, Katz-Vafa,
  mt-singularities, exotic}.
We restrict attention in this paper to standard matter representations
that can be realized through local enhancements of the Kodaira
singularity type, and  consider only briefly cases with local singularities
in the elliptic fibration at codimension two where $(f, g)$ vanish to
orders at least $(4, 6)$; such singularities have no direct Calabi-Yau
resolution preserving a flat elliptic fibration
and are associated with the appearance of superconformal
sectors in the F-theory model
\cite{Seiberg-Witten,Morrison-Vafa-II,Heckman-Morrison-Vafa,DelZotto-Heckman-Tomasiello-Vafa}.

One aspect of the connection between F-theory geometry and the anomaly
structure of the associated 6D supergravity theory that we will find
useful is the connection between the anomaly coefficient $b$ and the
genus of the associated complex curve $C$ in the base $B$.  We define a
``genus'' $g$ through
\begin{equation}
 2 g - 2 = b \cdot b + a \cdot b
\label{eq:genus-formula}
\end{equation}
for any gauge factor with anomaly coefficient $b$ in a 6D supergravity
theory.  In the F-theory picture this is just the genus of the
corresponding complex curve $C$; this genus formula can be decomposed into
contributions from each distinct matter representation, where some
more exotic matter representations are associated with singularities
in $C$ contributing to the arithmetic genus of $C$ \cite{KPT, exotic}.

The connection between F-theory geometry and physics for 4D F-theory
constructions of supergravity theories is similar but has additional
subtleties.  The gauge group  of the 4D theory again has a  piece
arising from 7-branes associated with Kodaira singularities in the
elliptic fibration over codimension one loci (complex surfaces in
this case) in the base threefold $B$, but additional gauge factors can
arise from D3-branes in the type IIB picture.  Furthermore, fluxes can
break the geometric gauge group down to a smaller group, and the
presence of a superpotential can push the complex structure moduli of
the compactification to regions in moduli space with gauge
enhancement.  For further details concerning the structure of 4D theories that
arise from F-theory see \cite{Morrison-TASI, Denef-F-theory,
  WT-TASI, Grimm-Taylor, Weigand-TASI, whatF}.

\section{The completeness hypothesis and massless charge sufficiency}
\label{sec:hypotheses}

\subsection{Statement of conditions}

The charge completeness hypothesis
specialized to 4D and 6D supergravity
states that
\vspace*{0.1in}

\noindent
{\bf  Charge completeness hypothesis for  4D and 6D supergravity:}

{\it Consider any consistent 4D ${\cal N} = 1$ or  6D ${\cal N} = (1, 0)$ supergravity
  theory.  States exist with all possible values
  in the charge lattice of the gauge group of the theory.}
\vspace*{0.1in}

We state the condition of 6D massless charge sufficiency  as
follows:
\vspace*{0.1in}

\noindent
{\bf Massless charge sufficiency condition  in six dimensions:}

{\it Consider any 6D supergravity
  theory.  The  massless states in the theory with nontrivial
  charges under the gauge group generate a charge lattice $\Lambda$.
  We say that the theory satisfies the ``massless charge sufficiency''
  condition when $\Lambda$ is the full charge lattice of the theory as
  given by  the charge completeness hypothesis.
  Equivalently, the massless charge sufficiency condition states that
  the global structure of the gauge group is such that the group acts
  effectively on $\Lambda$, i.e., there is no element of the gauge
  group that acts trivially on all massless charged states.}
\vspace*{0.1in}

\subsection{Proof of charge completeness hypothesis in F-theory}
\label{sec:completeness-proof}

We start with the completeness hypothesis in 6D.  The charge
completeness hypothesis for the connected $U(1)^k$ group of a 5D
M-theory compactification on a smooth Calabi-Yau threefold $Y$ follows
immediately from the fact that Poincar\'{e} duality gives a dual
pairing between the geometric structures encoding the gauge bosons and
the charged particles \cite{LieF}.  In this situation the U(1) gauge
bosons $A_i$ arise from the M-theory 3-form $C=A_i\wedge \omega_i$
where $\omega_i$ is a class in $H^{1, 1}(Y,\Z)$.  States that are
charged under the U(1) factors come from M2-branes wrapped on
Poincar\'{e} dual 2-cycles in $H_{1, 1}(Y,\Z)$.  The story in the
corresponding F-theory is conceptually similar but slightly more
subtle since some of the divisors in the corresponding Calabi-Yau
threefolds are not associated with 6D gauge bosons and must be
projected out to make a precise statement.

Assume that a 6D
supergravity theory has gauge group $G$ and comes from an F-theory
Weierstrass model that defines an elliptic Calabi-Yau threefold $X$
with resolution $\tilde{X}$.  We prove a ``connected'' version of the
completeness hypothesis in which we only consider charges under the
connected component of the identity in the gauge group -- that is, we
ignore discrete charges (if any).  Our goal is to prove that the
completeness hypothesis for a 6D supergravity theory realized through
F-theory is equivalent to the condition that the global structure of
the gauge group is identified through the standard assumption that the
connected part of the gauge group is $G^0 = (G_0 \times U(1)^r)/\Xi$,
where $G_0$ is simply connected and
$\pi_1 (G^0)$ coincides with the Mordell--Weil group of
the elliptic fibration.  As discussed in \S\ref{sec:F-theory}, each
gauge boson in the Cartan subalgebra of $G$ is associated with a
divisor $D \in H^{1, 1}(\tilde{X})$.  In the M-theory picture, these
gauge bosons arise by reducing the M-theory three-form on the given
divisor.\footnote{Note that there are also non-Cartan gauge bosons,
  which are obtained in the M-theory picture as part of the spectrum
  determined by wrapping M2-branes on these same divisors \cite{WitMF}.}  All
massive or massless charged states are associated (in the M-theory
picture) with M2-branes wrapped on elements of $H_2 (\tilde{X})$.  By
Poincar\'{e} duality there is a complex curve $C$ in $H_{1, 1} (\tilde{X})$
with $D \cdot C = 1$ for each divisor $D \in H^{1, 1}(\tilde{X})$.
More generally, there is a dual pairing between the linear space of
divisors $D$ (with coefficients in $\Z$) and the space of complex curves $C$.
Not all divisors in $H^{1, 1}(\tilde{X})$ correspond to massless gauge
bosons in the 6D F-theory model, however.  From the Shioda-Tate-Wazir formula
\cite{MR1610977,MR0429918,stw}, we know that $h^{1, 1}(\tilde{X}) = {\rm rk}\ G + h^{1,1}
(B)+1$; the divisors not corresponding to 6D gauge bosons are associated
with the zero section $\sigma_\infty$ (which defines the additive
identity in each elliptic fiber) and the pullbacks $\pi^{*}C_\alpha$ of divisors in the
base, where $\{C_\alpha\}$ is a basis of divisors in
the base.  Thus, to apply Poincar\'{e} duality to get a dual pairing
between gauge bosons and matter curves, we must separate out these
extra divisors.  The intersection of the zero section with the curve
associated with the generic fiber $F$ is always $\sigma_\infty \cdot F
= 1$.  Thus, we can always project away from this divisor-curve pair,
considering only curves orthogonal to $\sigma_\infty$ and divisors
orthogonal to $F$.  The pullbacks $\pi^{*}C_\alpha$ are always
orthogonal to $F$, $\pi^{*}C_\alpha \cdot F = 0$.  We can find a set
of dual curves $\hat{C}_\alpha = \pi^{*}C_\alpha \cdot \sigma_\infty -
(K_B \cdot C_\alpha) F$ that satisfy $ \sigma_\infty \cdot
\hat{C}_\alpha = 0, \pi^{*}C_\alpha \cdot \hat{C}_\beta =
\Omega_{\alpha \beta}$, where $\Omega_{\alpha \beta}$ is the
intersection form on the base $B$, by projecting out the appropriate
multiplicity of $F$, using the fact that, for any complex curve $C$ on the base,
 $\sigma_\infty \cdot
\sigma_\infty \cdot \pi^{*}C = K_B \cdot C$ where $K_B$ is the
canonical class of the base.

Projecting out these divisor-curve dual pairs, we are left with a dual
pairing between the remaining divisors and curves.  (Note that since
$\Omega$ is unimodular for any surface, the divisors and curves
that are pulled back from $B$ can always be projected out leaving a
dual pairing between the remaining divisors and curves).  The
remaining divisors consist of projections of sections (corresponding
to the Mordell--Weil group), and projections of the Cartan divisors of
the Lie algebra $\mathfrak{g}$.  The projected sections take the form
$\sigma-\sigma_\infty- \pi^*K_B$, where
$\sigma-\sigma_\infty$ fill out the Mordell-Weil group
$\Z^r \times T$ where $T$ is the Mordell-Weil torsion.  The Cartan
divisors of the Lie algebra are associated with the exceptional components of the
Kodaira fibers of the codimension one singularities in the resolved
fibration.  Together, the Cartan divisors and the sections in the
Mordell-Weil group (including the torsion elements) thus give a lattice which is
dual to the charge lattice of the theory.  Charge completeness holds
precisely when these divisors generate the fundamental group of the
Cartan torus of the gauge group, where the curves lie in the dual
lattice and therefore give all weights in the dual charge lattice
$\Lambda$.  This establishes the assertion that the (connected)
completeness hypothesis  for this class of theories is equivalent
to the standard assumption that  the fundamental group $\pi_1
(G^0)$ of the connected component $G^0$ of the 6D gauge group $G$ is
generated by the Cartan divisors and the full Mordell-Weil group of sections of the Calabi-Yau
threefold $X$.
In particular, the
global form of the gauge group is
$G^0 = (G_0 \times U(1)^r)/\Xi$, where $G_0$ is simply connected and
$\Xi \cap G_0 = T$
is the Mordell-Weil torsion. (See the appendix for more details about this group.)

Essentially the same argument goes
through unchanged for 4D ${\cal N} = 1$ supergravity theories that
come from F-theory on elliptic Calabi-Yau fourfolds, for the part of
the gauge group that comes from Kodaira singularities over divisors in
the threefold base and from sections.  In 4D F-theory
models, gauge bosons in this part of the gauge group are associated with divisors in the same way as
in 6D theories.  There are again matter states associated with
M2-branes wrapped on complex curves in a similar way; note, however, that in
the 4D construction these curves are fibral curves contained in a
``matter surface'' that lies over a codimension two locus in the
base. Note also that, unlike in 6D, in 4D F-theory constructions other
connected gauge group factors can arise from, e.g., stacks of D3-branes, so the
proof does not necessarily extend to the full connected gauge group.
Note also that fluxes and the superpotential do not change this
conclusion.  While the superpotential can drive the theory to a locus
with a larger geometric gauge group, the dual pairing between divisors
and matter curves holds in the same fashion.  And while fluxes can
break the gauge group down from that naively determined by the
7-branes and Kodaira singularity structure, this simply removes part
of the gauge group; the dual pairing between the remaining gauge
components and matter charges must still hold.

\subsection{Massless charge sufficiency in 6D}

We conjecture that the massless charge sufficiency condition holds for the
lattice of charges in
each F-theory compactification giving a 6D supergravity theory,
for all the (nonabelian or abelian) gauge factors
(in the connected part of the gauge group) where the associated Green-Schwarz coefficient
$b$ (or $\tilde{b}$ for abelian factors)  satisfies the condition
\begin{equation}
- a \cdot b  >
0\,,
\label{eq:exceptions}
\end{equation}
where the inner product denotes the Dirac product on the string charge lattice.
  In F-theory this is equivalent to the condition that
each such gauge factor is associated with a divisor $C$ satisfying $- K
\cdot C > 0$, where $K$ is the canonical class of the base $B$ and
the inner product is the intersection form in the
homology lattice of the compactification space% 
\footnote{As mentioned in the previous section,
for nonabelian factors the divisor $C$ is the locus where the
seven-branes supporting the gauge factor are localized; for U(1)
factors the divisor is somewhat more obscure but in the simplest cases
can be associated with the divisor supporting an unHiggsed nonabelian
factor containing the U(1).}
$H_{1, 1}(B,\Z)$.

There are two specific
classes of cases where the positivity conditions
(\ref{eq:exceptions}) are violated and we are aware of theories that
violate the massless charge sufficiency condition:
\begin{itemize}
\item[1)]
  Cases  where
\begin{equation}
  b \cdot b
\leq -3 {\rm \ and\ } - a \cdot b = b \cdot b +2 \leq -1
\label{eq:exception-1}
\end{equation}
These correspond to situations where the negative intersection $- K
\cdot C <0$  gives rise
to a  ``non-Higgsable'' gauge factor \cite{clusters}, with 
minimal matter content (or possibly none at all), since the Weierstrass coefficients $f, g$ are
forced to vanish along the geometric locus $C$
associated with $b$.
\item[2)]
  Cases where
  \begin{equation}
b \cdot b= - a \cdot b = 0
\label{eq:exception-2}
  \end{equation}
 These cases are associated with gauge factors for which the
matter content contains precisely one hypermultiplet in the adjoint
representation and no other charged matter, as would be seen in a
multiplet with enhanced supersymmetry.
\end{itemize}
We explore these two classes of exceptions further in
\S\ref{sec:special-cases}.  In fact, our conjecture is equivalent to
the statement that massless charge sufficiency holds for all factors
except those that fit into these two exceptional classes.
This can be
seen as follows:
If the curve $C$ in the base associated with the anomaly coefficient $b$ is
irreducible and $- K \cdot  C <0$ it follows that%
\footnote{This can be shown for example by noting
  that such a negative intersection product is only possible for
  irreducible $C$ when $-K$ contains $C$ as a component; it follows
  that we can write $- K = n C + R$ with $R \cdot C \geq 0$ (Zariski 
decomposition \cite{MR0141668,MR2877664}), so $C \cdot C
  <0$.}
 $ C \cdot C =b \cdot b <0$.
Since the genus $g$ in (\ref{eq:genus-formula}) is non-negative, we
can only have a gauge factor associated with an irreducible curve $C$
and corresponding anomaly coefficient $b$ with $- a \cdot b<0$
when $g = 0$ so $ - a \cdot b = b \cdot b +2 $ as in case (1).  If $K
\cdot C = 0$ then the genus $g =1$ curve with $C \cdot C = 0$
corresponds to the situation of case (2) giving our other class of
exceptions.  When $K \cdot C = 0$ and the genus is higher, $g >1$,
then $- a \cdot b = 0, b \cdot b >0$, or $g = 0$ so $b \cdot b = -2$,
and it seems that massless charge
sufficiency always holds, as we see in a number of cases in the
following analysis.

While we do not have a mathematical
proof of this conjecture, it holds in all cases we have
analyzed, as we show in the following sections explicitly for various
single-factor gauge groups.  Here, we elaborate
slightly further on the statement of this conjecture.

As described above,
the connected part $G^0$ of the F-theory
gauge group $G$ is generally a compact group of the form
 $G^0 = (G_0 \times U(1)^r)/\Xi$, where $G_0$ is simply
connected and $\Xi$ is a discrete subgroup of the 
center of $G_0 \times U(1)^r$.
We can  consider the charge lattice $\Lambda_0$ of the group $G_0
\times U(1)^r$.
In any F-theory model, the massless states are associated with
holomorphic or anti-holomorphic fibral
curves (i.e.\ curves that project to a point in the base).
The associated massless hypermultiplet states generate a
sublattice $\Lambda \subset \Lambda_0$.
Unless there are U(1) factors under which no fields are charged
(``non-Higgsable U(1) factors'' \cite{Martini-WT, Wang-u1, MPT}), the
rank of $\Lambda$ is the same as the rank of $\Lambda_0$.
Projecting out the charge lattice associated with such U(1) factors
and with nonabelian factors
having anomaly coefficients that violate (\ref{eq:exceptions}),
to give
the lattice $p (\Lambda_0)$, and similarly projecting out the charges under
nonabelian factors violating (\ref{eq:exceptions}) from $\Lambda$,
the quotient $\Xi' = p(\Lambda_0)/p(\Lambda)$ is
the subgroup of the center of $G_0$ that leaves all the remaining
massless charged states invariant.
Our conjecture is then that $\Xi= \Xi'$.

In the next section we analyze the
possible Mordell-Weil torsion groups $T$ for nonabelian gauge
groups.  We show that for many of the
gauge factors that contain each such group
 $\Xi = T$ in the center, when the anomaly coefficient satisfies (\ref{eq:exceptions})
it follows that
massless charge sufficiency always holds. In particular, under these circumstances
all anomaly-consistent massless spectra that
 are invariant under $\Xi = T$ are associated with an  F-theory
 Weierstrass model that indeed has this torsion
group, so that $\Xi= \Xi'$ in all of these cases.  In Section
\ref{sec:abelian} we give an example of massless charge sufficiency
for an abelian U(1) theory, and in Section \ref{sec:special-cases}, we
describe in more detail the cases
that violate (\ref{eq:exceptions})
and the massless charge sufficiency condition

\section{Examples: massless charge sufficiency and Mordell-Weil
  torsion for nonabelian gauge groups}
\label{sec:nonabelian}

We present here a variety of examples of 6D supergravity theories with
purely nonabelian gauge groups
that
are anomaly free and have a massless spectrum that is invariant under
a nontrivial discrete subgroup of the gauge group.
These are the situations in which the massless charge sufficiency
condition has the potential to fail.
We show that in
all these cases except those where the associated anomaly coefficient
$b$ satisfies (\ref{eq:exception-1}) or (\ref{eq:exception-2})
the Mordell-Weil group of any F-theory construction with the desired
gauge algebra and matter content has the appropriate torsion
subgroup $T = \Xi$, where $G^0 = G_0/\Xi$, so that the massless charge  sufficiency condition is
satisfied, in accord with our conjecture.  
We focus primarily here on theories with ``generic'' matter
representations \cite{TT-generic}, which for a single $SU(N)$ factor,
for example, consists only of the fundamental, adjoint and two-index
antisymmetric representations. We explicitly analyze  one case of exotic
(non-generic) matter in \S\ref{sec:z3} (the 3-index antisymmetric
representation of SU(6)). The analysis of \cite{exotic} suggests that
there are few other viable exotic matter representations compatible
with Mordell-Weil torsion; in particular, the 3-index ({\bf 4})
representation of SU(2) breaks the possible $\Z_2$ central subgroup,
while the 2-index symmetric representation of a general $SU(N)$ group
can preserve the $\Z_2$ subgroup of $\Z_N$ for even $N$.
In general, including more exotic matter types does not seem to modify
the conclusions found here; we make a few comments about this in
various appropriate places but do not make an effort towards a
complete analysis of such questions.
In this section and the following sections we use some basic aspects
of the geometry of elliptic curves; for a good introduction to these
topics and related methods see \cite{Silverman-Tate, Silverman}.

We organize the analysis by the Mordell-Weil torsion group, and
consider for each torsion group different gauge groups that may arise
with that torsion subgroup.  While our analysis here is not completely
exhaustive, we give a fairly complete analysis of the
possibilities for nonabelian gauge groups.

From the geometry of elliptic Calabi-Yau manifolds we know that the
only possible torsion groups that can appear in the Mordell-Weil group
for an elliptic Calabi-Yau threefold or fourfold are the following
groups:\footnote{We use additive notation for these groups,
which are always abelian.}
\begin{equation}
  \Z_m, 2 \leq m \leq 6; \; \;
  \Z_2 \oplus\Z_{2n}, 1 \leq n \leq 2; \; \;
  \Z_3 \oplus\Z_3 \,.
 \label{eq:possibilities}
\end{equation}
Explicit forms of the Weierstrass models associated with each of these
discrete groups were given in \cite{Aspinwall-Morrison}, following
\cite{Kubert}.  It was argued in \cite{Hajouji-Oehlmann} that
these are the only possible torsion groups for elliptic Calabi-Yau
manifolds.\footnote{Note that, as shown by Mazur \cite{Mazur}, other groups such
as $\Z_7, \ldots,\Z_{10},\Z_{12}$ and $\Z_2\oplus\Z_6, \Z_2\oplus\Z_8$ 
are possible for  an elliptic
curve over $\Q$, though $\Z_3 \oplus \Z_3$ is not allowed over $\Q$.}  We consider in turn the Weierstrass forms and 6D supergravity
models associated with theories with the torsion groups (\ref{eq:possibilities}).

On the 6D supergravity side, we consider models with different
nonabelian gauge groups.  Abelian gauge factors are more subtle and we
address some questions about U(1) factors in the following section.
For nonabelian gauge algebras that can appear in 6D supergravity
theories, the discrete central subgroup of the simply connected
associated gauge group is listed in Table~\ref{t:discrete-centers},
following \cite{Bourbaki}.
\begin{table}
  \begin{center}
    \begin{tabular}{ | c | c | c | }
      \hline
  gauge algebra & simply connected $G_0$ & center\\
  \hline
  $  \gsu(n) $& SU($n$) & $\Z_n$\\
  $  \gsp(n) $& Sp($n$) & $\Z_2$\\
  $\gso(4n)$ & spin($4n$) & $\Z_2 \oplus \Z_2$\\
  $\gso(4n +2)$ & spin($4n +2$) & $\Z_4$\\
  $\gso(2n +1)$ & spin($2n +1$) & $\Z_2$\\
  $\age_6$ & $E_6$ & $\Z_3$\\
  $\age_7$ & $E_7$ & $\Z_2$\\
  $\age_8$ & $E_8$ & 1\\
  $\agf_4$ & $F_4$ & 1\\
  $\gggg_2$ & $ G_2$ & 1\\
  \hline
\end{tabular}
\end{center}
\caption[x]{\footnotesize  The discrete central subgroup of the simply
connected groups associated with each gauge algebra}
\label{t:discrete-centers}
\end{table}

\subsection{$\Z_2$ torsion}

\subsubsection{General Weierstrass models with $\Z_2$ torsion}

A Weierstrass model that has a section of order 2, corresponding to a
$\Z_2$ factor in the Mordell-Weil group, must have the form
\cite{Aspinwall-Morrison}
\begin{equation}
y^2 = x (x^2 + \alpha_2x + \alpha_4) \,.
\label{eq:Weierstrass-2}
\end{equation}
This can be seen simply from the elliptic curve addition law, which
states that for a given elliptic curve $C$ the three points $p, q, r$
that lie on the intersection $L \cdot C$ with any line $L$ satisfy $p +
q + r = 0$, and the condition that the zero point ${\cal O}$ of the
elliptic curve lies at the point at infinity along any vertical line
$x = \rm{const}$ in the $x, y$ plane.  
From these conditions, we see that a point $p$
satisfies $2 p = p + p = 0$ if and only if the tangent to the cubic at
$p$ is vertical, so shifting the point $p$ to the origin gives the
Weierstrass form (\ref{eq:Weierstrass-2})

Completing the square to get the reduced form we have
\begin{equation}
y^2 = x^3 + fx + g
% \label{eq:}
\end{equation}
where
\begin{align}
f  &=-\alpha_2^2/3 +  \alpha_4\nonumber\\
g & = 2 \alpha_2^3/27-\alpha_2 \alpha_4/3 \,.
\label{eq:z2}
\end{align}
The discriminant then has a factor
\begin{equation}
\Delta = 4f^3 + 27g^2 = \alpha_4^2 \tilde{\Delta} \,,
% \label{eq:}
\end{equation}
implying at least a gauge algebra $\gsu(2)$ on the locus $\alpha_4 = 0$.

\subsubsection{$\gsu(2)$ Weierstrass models with $\Z_2$ torsion}
\label{sec:su2}

We begin with models with a gauge algebra $\gsu(2)=\gsp(1)$.
The generic form of Weierstrass model giving a gauge algebra $\gsu(2)$
(which generically gives an SU(2) group that has only fundamental and
adjoint matter) on a curve $\sigma = 0$ can be constructed from the
Tate form \cite{Bershadsky-all, Katz-etal-Tate}, and by direct
analysis of the general Weierstrass model with the appropriate Kodaira
singularity type \cite{mt-singularities}; from the latter perspective
it is clear that this Weierstrass model is the most general of this
type and has all the moduli associated with this branch of 6D
supergravity theories.  Such a generic SU(2) Weierstrass model has
\begin{align}
 f & = -\frac{1}{48} \phi^2 + \sigma f_1\nonumber\\ g &
 =\frac{1}{864} \phi^3 -\frac{1}{12} \phi \sigma f_1 + \sigma^2
 ({g}_2) \,,
\label{eq:su2}
\end{align}
which can be related to the Tate parameters $a_k=\sigma^{n_k} \tilde{a}_k$ 
(with $n_1, n_2, n_3, n_4, n_6 = 0, 0, 1, 1, 2$)
by
\begin{equation}
f = -\frac{1}{48}  (a_1^2 + 4a_2)^2 + (\tilde{a}_4 + a_1 \tilde{a}_3/2) \sigma \,,
% \label{eq:}
\end{equation}
and
\begin{equation}
g =\frac{1}{864}   (a_1^2 + 4a_2)^3  -\frac{1}{12} 
(a_1^2 + 4a_2)
(\tilde{a}_4 + a_1 \tilde{a}_3/2) \sigma + (\tilde{a}_6 + \tilde{a}_3^2/4) \sigma^2 \,.
% \label{eq:}
\end{equation}
We thus identify $\phi = a_1^2 + 4a_2$, $f_1 = \tilde{a}_4 + a_1
\tilde{a}_3/2$, $g_2= \tilde{a}_6 + \tilde{a}_3^2/4$, and we see
that the Tate formulation has a redundancy under the choice of $a_1,
a_3$.
Matter in
the fundamental representation of $\gsu(2)$  transforms nontrivially
under
 the $\Z_2$ center,
while matter in the adjoint representation does not, so the
possibility of having a group of global structure $SO(3) = SU(2)/\Z_2$
can only arise in the absence of matter in the fundamental
representation.
% WT added preceding sentence for clarity since it wasn't stated elsewhere.

The discriminant locus of (\ref{eq:su2}) takes the form
\begin{equation}
\Delta = \frac{1}{16}  \phi^2 (\phi g_2-f_1^2)  \sigma^2 +{\cal O} (\sigma^3)\,.
% \label{eq:}
\end{equation}
Fundamental matter fields arise at the vanishing locus of the factor
$(\phi g_2-f_1^2)$.   In general, $\phi$ is a section of ${\cal O}
(-2K)$, and $\Delta$ is a section of ${\cal O} (-12K)$.

It is straightforward to confirm that when $g_2 = 0$, the Weierstrass
model (\ref{eq:su2}) takes the form (\ref{eq:z2}) and has $\Z_2$
torsion.  We now prove that this is the case whenever $- a \cdot b >0$
and there is no fundamental matter, only matter in the adjoint of
$\gsu(2)$.
From the anomaly cancellation conditions it is straightforward to
determine that the number of hypermultiplet fields in the fundamental
representation of a theory with $\gsu(2)$ algebra is $- 8 a \cdot b -
2 b \cdot b$ (assuming as stated above that we have only generic
matter types, so that the only representation arising is adjoint
matter).  To have no fundamental matter fields, we must then have the
condition
\begin{equation}
 (-4 a - b) \cdot b = 0 \,.
% \label{eq:}
\end{equation}
If $- a \cdot b >0$, we must then also have $b \cdot b = -4 a \cdot b
>0$.
The class of $g_2$ is $[g_2] = -6 a - 2 b$.  It thus follows that
$[g_2] \cdot b = - b \cdot b/2 <0$.  This, however, implies that
$[g_2]$ is not an effective class, since if $b$ is an irreducible
effective class with $b \cdot b >0$ there is no effective class $c$
with $c \cdot b <0$.
This shows that $g_2$ must vanish, so the absence of fundamental
matter implies the presence of the
 Mordell-Weil torsion section associated with the
$\Z_2$ quotient of the gauge group.  This demonstrates that, at least
for theories with only the generic fundamental and adjoint matter
representations of a gauge factor $\gsu(2)$, when the condition $- a
\cdot b >$ 0 is satisfied the massless charge sufficiency condition
holds.  Note that for the gauge algebra $\gsu(2)$, the only exceptions
to massless charge sufficiency are those of class (2) where the
condition (\ref{eq:exception-2}) holds.  When $b \cdot b \leq - 3$,
the rank of the resulting non-Higgsable gauge algebra is at least
two.  And when $- a \cdot b = 0, b \cdot b = -2$, we have fields in
the fundamental matter representation so the charge lattice is
automatically filled.

For the gauge algebra $\gsu(2)$, there cannot be any further
exceptions to massless charge sufficiency even when other matter
representations that can be realized through F-theory geometry are
included.  It is argued in \cite{exotic} that the only exotic matter
representation that can arise in F-theory for $\gsu(2)$ is the
3-index ({\bf 4}) representation.  Since this representation by itself
is sufficient to generate the full SU(2) charge lattice, any model
that includes this representation automatically satisfies massless
charge sufficiency.

It is interesting to consider explicitly some of the classes of
cases with generic matter
where  the massless charge sufficiency condition holds.
When $\sigma$ is in the class $-4K$, $f_1$ becomes a constant and
$g_2$ automatically vanishes.  In this case,  as discussed above,
there is no (massless)
matter charged in the fundamental representation of the $\gsu(2)$,
only adjoint matter.
The presence of the $\Z_2$ torsion in this case indicates that the
global form of the gauge group is SO(3) $ = SU(2)/\Z_2$, in agreement
with the massless charge sufficiency hypothesis.

A special case of the $SU(2)/\Z_2$ Weierstrass model occurs when $\phi
= 0$ and $[\sigma] = -4K_B$.  In this case, $g$ vanishes identically
and we have a Kodaira type III singularity, giving an alternative
realization of the $\gsu(2)$ algebra.

When $\sigma$ is in a class such that $[g_2] = -6K_B-2[\sigma]$ is
ineffective (there are no sections of the corresponding line bundle),
then $g_2$ automatically vanishes.  In such a case, as long as $f_1$
is in an effective class (has sections), we get another SU(2) factor
on the locus $f_1 = 0$ (or several SU(2) factors if $f_1$ is
reducible).\footnote{This is a simple example of the automatic
  enhancement conjecture described in \cite{Raghuram:2020vxm}, and was
  part of the motivation for the work in that paper.}  This can also occur when $g_2$ is in an effective class
but tuned to vanish.  In either of these cases, we again have $\Z_2$
torsion, but the global form of
the gauge group is $G = (SU(2) \times
SU(2))/\Z_2$ (with more factors if $f_1$ or $\sigma$ is reducible).
(Note that if
$f_1$ is not in an effective class, or otherwise vanishes along with
$g_2$, then $f, g$ vanish to order (2, 3) at the vanishing locus of
$\phi$, which supports a larger gauge group such as SO(8)).

As simple examples of these situations, consider the base $\P^2$, with
canonical class $-K_B = 3H$.  When $\sigma$ is a degree 12 curve, we
have an SO(3) theory with only adjoint matter (55 adjoints).  When
$[\sigma] < 12$ and
$g_2 = 0$ we have an $(SU(2) \times SU(2))/\Z_2$ theory where the
divisors supporting the two $SU(2)$ factors have degrees $d, 12-d$.
When the degree of $\sigma$ is 10 or 11, this condition is forced
automatically, so there is no pure SU(2) theory with anomaly
coefficient $b = d = 10, 11$ that comes from such an F-theory model at
$T = 0$.  Such theories are anomaly-consistent, however, so form a
part of the apparent 6D supergravity ``swampland''.

Note that the cases with gauge group $(SU(2) \times SU(2))/\Z_2$ can
be thought of as degenerate cases of the SO(3) theory where the
divisor supporting the gauge group becomes reducible.  In particular,
when $[\sigma] +[f_1] = -4K$, this can be thought of as tuning an
SO(3) on a reducible divisor $\alpha_4 = \sigma f_1$.

From the 6D supergravity point of view, the models described correspond
precisely with the set of anomaly-consistent models with  only SU(2)
gauge group factors and generic matter content that is invariant under
a central $\Z_2$.  For example, at $T = 0$ the number of fundamental
matter representations of an SU(2) with anomaly coefficient $b > 0$ is
easily computed to be $2 b (12-b)$.
This vanishes only when $b = 12$, corresponding to the SO(3) theory
described above, and the fundamentals are all associated with
bifundamentals  with other SU(2) factors precisely when the anomaly
coefficients of the SU(2) factors satisfy $\sum_{i}b_i = 12$.

\subsubsection{$Z_2$ torsion and larger $SU(N)$ and $Sp(N)$ groups}
\label{sec:sun}

For groups $SU(2N)$  with $N > 1$, the discrete $\Z_2$ subgroup of the
center $\Z_{2 N}$ is respected by models that have no fundamental
matter but do have matter in the 2-index antisymmetric and adjoint
representations. 
The $\Z_2$ center of $Sp(N)$ models without fundamental matter is
similarly preserved, and the anomaly equation for the number of
fundamental matter fields is the same in these cases.
A variety of such models with larger $SU(N)$ and $Sp(N)$
groups can be constructed by
taking $\alpha_4 = \sigma^k$ in (\ref{eq:z2}) for some integer $k$.  In this case the
gauge group on the locus $\sigma = 0$ becomes $SU(2k)$ when $\alpha_2$
is a perfect square, and otherwise is $Sp(k)$.  Since $\alpha_4$ is in the class $-4K$, 
the divisor $4K$ must be divisible by $k$ in $\operatorname{Pic}(B)$ in order
to carry out 
such a construction.  For example, on the base
$\P^2$ we have such a model for any $k | 12$, so that there are 
$T = 0$
models
with gauge groups $SO(3), SU(4)/\Z_2, SU(6)/\Z_2, SU(8)/\Z_2,
SU(12)/\Z_2$, and $SU(24)/\Z_2$  (and similarly for the
analogous $Sp(N)/\Z_2$ models).  In each case there are
no fundamental matter fields, only adjoint and two-index antisymmetric
tensor fields.  The Weierstrass model for $SU(24)/\Z_2$ was considered
previously in \cite{mt-singularities} (\S 4.1).
It is not hard to check that these are the only anomaly-free models
with $T = 0$ and gauge algebra $\gsu(N)$ or $\gsp(N)$ with generic
matter representations
and no matter in the
fundamental representation.
Note that there may be related models (in some cases connected by
``matter transitions'' \cite{matter-transition}) in which there are
two-index symmetric and antisymmetric
matter representations replacing adjoints \cite{exotic}, which
could still preserve the $\Z_2$ subgroup; we leave further analysis of
such possible exotic matter models  as open problems.
Other models with a
variety of product gauge groups can be
constructed by making $\alpha_4$ reducible.  For example, taking $T =
0$ and $\alpha_4 = \phi_4^2 \phi_2^2 $ gives a theory with gauge group
$(SU(4) \times SU(4))/\Z_2$ on degree 4 and 2 curves in $\P^2$, so
that the matter content consists of 3 adjoint representations of the
first SU(4), 8 bifundamental matter fields, and 12 and 6 two-index
antisymmetric matter fields for the two SU(4) factors respectively.

Note that making $\alpha_4$ reducible in the special case where
$\alpha_2$ = 0 enhances the Kodaira type III singularity mentioned
above to a type $I_0^*$ (SO(8)) or type $III^*$ ($E_7$) singularity
when $\alpha_4 = \sigma^2$ or $\alpha_4 = \sigma^3$ respectively, as
discussed further below.

Considering  $SU(N)$ models with $T > 0$, 
 a
  similar, though slightly more intricate, analysis to that of SU(2)
allows us to prove that just as for SU(2), for any SU(4) model (for
any $T$) where $-a \cdot b > 0$ that has only generic matter invariant
  under the $\Z_2$ subgroup of the center the associated Weierstrass
  model has  $\Z_2$ torsion.  This can be seen as follows: from
  \cite{mt-singularities}, the general form of a Weierstrass model
  with SU(4) symmetry on the divisor $\sigma$ is
\begin{align}
f & =  -\frac1{48} \phit^4 -\frac16 \phit^2 \phif \sigma + f_2 \sigma^2 + 
f_3\sigma^3  \label{eq:4-Weierstrass}\\
g & =   \frac1{864} \phit^6+\frac1{72} \phit^4 \phif \sigma
+ (\frac1{36} \phit^2 \phif^2 -\frac1{12}\phit^2 f_2) \sigma^2 +
( -\frac1{12}\phit^2 f_3 -\frac13  \phif f_2 -\frac1{27} \phif^3)
\sigma^3 %\\&\quad
 + {g}_4 \sigma^4\,,
 \notag 
\nonumber
\end{align}
where $f_3, g_4$ may also include terms of higher order in $\sigma$.
From anomaly cancellation for a theory with only generic (adjoint,
two-index antisymmetric, fundamental) matter, the number of fields in
the fundamental representation is $-8a \cdot b-4b \cdot b$, so to have
no matter fields we must have
\begin{equation}
(-2a-b) \cdot b = 0 \,.
% \label{eq:}
\end{equation}
If $-a \cdot b > 0$, it follows that $b \cdot b > 0$.  The class of
    $f_3$ is $[f_3] = -4a-3b$, which must be ineffective.  Similarly,
    $g_4$ is ineffective, so $f_3 = g_4 = 0$.  The remaining nonzero terms in the
    Weierstrass model (\ref{eq:4-Weierstrass}) then take the form
    (\ref{eq:z2}), where
\begin{align}
 \alpha_2 & = \frac{1}{4} \phi_0^2 + \phi_1 \sigma\\
\alpha_4 & = (f_2 +\frac{1}{3} \phi_1^2) \sigma^2
\end{align}
This shows that our conjecture on massless charge sufficiency holds
for SU(4) models with only two-index antisymmetric and adjoint
matter.   The only other representation possible in standard
F-theory models is the two-index symmetric representation
\cite{exotic}, and the adjoint is anomaly equivalent to a combination
of the two-index symmetric and antisymmetric representations, so the
condition for the number of fundamental representations to vanish is
the same in the presence of this exotic matter.  Thus, for SU(4) we
have proven the massless charge sufficiency conjecture for arbitrary
matter fields.
The same argument holds for models with a gauge group Sp(2); the only
difference is that we replace the perfect square $\phi_0^2$ (that
satisfies the ``split'' condition) in the Weierstrass model with a
more general field $\phi$.

A similar argument holds for $SU(2 k)$ models with larger $N = 2k$.  While
for SU(6) and SU(8) the story is slightly more complicated due to the
possible presence of 3-index antisymmetric matter representations, the
most generic type of $SU(2 k)$ tuning takes the form
  \cite{mt-singularities}
\begin{align}
f & =   -\frac13 \Phi^2 + f_k \sigma^k + f_{k + 1} \sigma^{k + 1}
\label{eq:expansion-f-even}\\
g  & =   -\frac{2}{27} \Phi^3  - \frac13 \Phi (f_k \sigma^k + f_{k +
  1} \sigma^{k + 1}) + g_{2k}\sigma^{2k} \,.
\nonumber%\label{eq:expansion-g-even}\,
\end{align}
The number of fundamental matter fields is given by $-8a \cdot b-2k\ b
\cdot b$, so to have no fundamental matter fields we must have
\begin{equation}
(-4a-kb) \cdot b = 0 \,.
% \label{eq:}
\end{equation}
The classes of $f_{k + 1}$ and $g_{2k}$ are $-4a-(k + 1)b$ and $-6a-
2k$, which are both ineffective through similar arguments to the
above, so these terms vanish and the absence of fundamental matter
with $-a \cdot b$ thus implies that we have the $\Z_2$ torsion form
(\ref{eq:z2}).
Again, the analysis is parallel for $Sp(k)$.
All these models thus agree with the conjecture for massless charge sufficiency.

\subsubsection{$E_7$ gauge group}

The group $E_7$ has a center $\Z_2$.  The generic matter
representations are the adjoint ({\bf 133}) and fundamental ({\bf 56})
representations.  No higher-dimensional exotic $E_7$ representations
should be possible in standard F-theory models.
The adjoint is invariant under the $\Z_2$ center, while the
fundamental is not.  Thus, the only situation where we can have a
matter spectrum with generic matter that respects the $\Z_2$ symmetry
is when we only have adjoint matter.  We get an $\age_7$ algebra over a
divisor $\sigma = 0$ for a Weierstrass model with $f, g$ of the form
\begin{align}
  f & = \sigma^3 \tilde{f}\\
  g & = \sigma^5 \tilde{g} \nonumber\\
  \Delta & =  4 \sigma^9 \tilde{f}^3 \nonumber
% \label{eq:}
\end{align}
Fundamental matter appears at the locus where $\tilde{f}= 0$.  

We now prove that the massless charge sufficiency conjecture holds by
demonstrating that the absence of fundamental matter and the condition
$-a \cdot b$ together imply the presence of $\Z_2$ torsion.  The number of
(half-hypermultiplet) 
fundamental matter fields is given
by $(-4a-3b) \cdot b$.  When this quantity vanishes, it follows that
$\tilde{g}$, with class $[\tilde{g}]= -6a-5b$ has an ineffective class
and vanishes.
We thus match the Weierstrass form
(\ref{eq:Weierstrass-2}) with $\alpha_2 = 0, \alpha_4 =
\tilde{f}\sigma^3$ when the $E_7$ theory has only adjoint
matter, giving another class of examples where the massless charge sufficiency
conjecture holds.

In the simplest cases with only adjoint matter $\tilde{f}$ is simply a
constant, and $\sigma$ is a section of $-(4/3)K_B$.  This is possible,
for example, on the base $\P^2$ when $\sigma$ is a quartic polynomial.
Other models with $\Z_2$ torsion can arise when $T > 0$.

One other relevant class of models occurs when $\tilde{g} = 0$ but
$\tilde{f}$ is not a constant.  This corresponds to the case discussed
in the previous subsections where $\alpha_2= 0$ and $\alpha_4$ is reducible,
where we have a product group with both an $E_7$ and other factors.

\subsubsection{$SO(N)$ and $\Z_2$ torsion}
\label{sec:so}

A theory with an $\gso(N)$ gauge algebra and generic matter generally
has both fundamental matter and matter in the spinor representation,
in addition to adjoint matter.  We focus here first on the cases $\gso(2n +
1)$, where the center of the simply connected group spin$(2n + 1)$ is
$\Z_2$.

The Weierstrass model for an $\gso(2n + 7)$ 
or $\gso(2n + 8)$
gauge algebra has the form
\begin{align}
  f & =  f_2\sigma^2 + f_3 \sigma^3 \label{eq:so}\\
  g & = g_3\sigma^3 + g_4 \sigma^4  \nonumber
\\
 \Delta & =  \Delta_{6 + n} \sigma^{6 + n}\,, \nonumber
\end{align}
where the choice of $2n + 7$ or $2n + 8$ depends upon additional
monodromy conditions, and the last coefficients in each expansion may contain
higher order powers in $\sigma$.

The spin representation breaks the central $\Z_2$ for odd $N$, so we
can have $\Z_2$ torsion with gauge algebra $\gso(2n + 1)$ only when there
are no fields in the spin representation.  By anomalies this occurs
only when $(-2a-b) \cdot b = 0$. In this circumstance, when $-a \cdot
b > 0$ it follows that $f_3, g_4$ vanish.  The discriminant is then
  $\Delta = (4f_2^2 + 27g_3^{3}) \sigma^6$, so this is only possible
  for $\gso(7)$ when $N$ is odd.  The $\gso(7)$ monodromy condition is
  that the cubic $x^3 + f_2x + g_3$ factorizes as $(x-A) (x^2 + A x +
  B)$, while a factorization of the form $(x-A) (x-B) (x + (A + B))$
  gives $\gso(8)$.  When $(-2a-b) \cdot b = 0$ and $-a \cdot b > 0$ we
    have $b \cdot b > 0$, and it follows that $-2a-b$ is a rigid
      divisor (or sum of rigid divisors).\footnote{We argue in favor of
this by using the Zariski decomposition \cite{MR0141668,MR2877664}
of the effective $\Q$-divisor $-2a$.  (This divisor is effective
because $-2a=-2K_B$ and the Weierstrass coefficients $f$ and $g$
are sections of $-4K_B$ and $-6K_B$, respectively.)  We assume that
the gauge divisor $C$ is irreducible; since $C\cdot C = b\cdot b > 0$,
it follows that $b$ is nef (since its intersection with any irreducible
divisor other than $C$ must be nonnegative).  Thus, $-2a = b + (-2a-b)$
must be the Zariski decomposition of $-2a$, so that in particular, if
we write $-2a-b$ in terms of irreducible components, the intersection
matrix of those components must be negative definite.  In particular,
each component has negative self-intersection and so it is rigid.}
It then follows that $f_2 = c A^2, g_3 = c'
      A^3$ with $c, c'$ constant and $[A] = -2a-b$.  This implies that
      the monodromy is that of $\gso(8)$.  In this case in fact the
      torsion is $\Z_2 \oplus \Z_2$, as we discuss further in
      \S\ref{sec:z22}.  This proves that the massless charge
      sufficiency conjecture we have made holds for gauge algebras
      $\gso(2n + 1)$ taken in isolation.  In fact,
we have shown that there are no
      situations at all where a pure $\gso(2n + 1)$ gauge algebra
      arises without some matter in the spinor representation when $-a
      \cdot b > 0$.

We could also get a $\Z_2$ torsion by breaking part of the discrete
center of the gauge group when the algebra is $\gso(2n)$ with $2n$
even.  For the algebra $\gso(4n + 2)$,  there are two spinor
representations and either spin representation again
completely breaks the central $\Z_4$ \cite{Bourbaki}.  From the same argument as
above, the absence of matter in a spinor representation implies $(-2a-b)
\cdot b = 0$, and
there are no models with $-a \cdot b > 0$ and no spinors except
  $\gso(8)$.
For the gauge algebra $\gso(N)$ with $N = 4n$, the presence of either
of the spinor
representations or the fundamental representation leaves a $\Z_2$
invariant, but the center is completely broken by any pair of these
representations.  
When the genus $g$ defined through (\ref{eq:genus-formula}) is one,
then the multiplicities of both spinor and fundamental representations
are given by $b \cdot b$, so the center is completely broken except in
the exceptional case  $-a \cdot b = b \cdot b = 0$.  When $g = 0$, the
number of spinors is $b \cdot b + 4$, which can only vanish when $b
\cdot b = -4$, another exceptional case that we return to in
Section \ref{sec:exceptional-non-Higgsable}, and the number of
fundamental representations is $b \cdot b + N -4$, which is always
larger than or equal to the number of spinors, so the only cases where the
discrete center is not completely broken are exceptional.  When $g  > 1$,
on the other hand, there are always more spinors than fundamental
representations when $N > 8$; the case of $\gso(8)$ is treated in
  \S\ref{sec:z22}.
This leaves as a possible class of exceptions to massless charge
sufficiency theories with algebra $\gso(4n)$ and spinor matter but no
fundamental matter.  For example, on the base $\P^2$ we can tune a
$\gso(12)$ algebra on a quartic (genus $g = 3$), and the matter
content consists of 8 spinor fields and no fundamental fields.   If
all the spinors appearing here are in the same representation, this
would violate massless charge sufficiency unless the corresponding
Weierstrass model has $\Z_2$ Mordell-Weil torsion.  We can check
explicitly, however, that in such a case the Weierstrass model does
have the form (\ref{eq:z2}).  For the number of fundamental fields to
vanish for $\gso(12)$, we have $(-4a-3b) \cdot b = 0$.  From this it
follows that the Tate coefficients $a_1, a_3, a_6$ must vanish in the
standard $\gso(12)$ Tate tuning, giving precisely the resulting
Weierstrass model form (\ref{eq:z2}).  It is not possible to have an
$\gso(16)$ or higher gauge algebra on a curve with $b \cdot b > 0$, so
we have checked massless charge sufficiency for all possible $\gso(n)$
algebras with $\Z_2$ torsion.

Considering product groups associated with gauge algebras with
multiple components, it would seem that various combinations of
$\gsu(2), \gso(7),$ and $\age_7$ algebras can be produced with a
single $\Z_2$ torsion factor by taking the class of $\Z_2$ models with
$\alpha_2 = 0$ and setting $\alpha_4$ to be a product of various
linear, quadratic, and cubic factors.  The $\gso(7)$ part of the
Weierstrass model then takes the form of (\ref{eq:so}), with $f_3 = g
= 0$.  While taken alone the $\gso(8)$ symmetry coming from the
quadratic factors would respect a $\Z_2 \oplus \Z_2$ torsion as
described below, the monodromy condition is changed to give an
$\gso(7)$ and the discrete symmetry is broken to $\Z_2$ in the
presence of at least one linear or cubic factor in $\alpha_4$.  When
the only gauge factors are $SU(2)$ and a single $SO(7)$, this gives a
perfectly well-behaved set of models where all matter is in ({\bf 2},
{\bf 8}) representations at the intersections of the $SU(2)$ divisors
with the $SO(7)$ divisor; this matter is invariant under the $\Z_2$
that acts simultaneously on all factors.  Including more than one
$SO(7)$ or $E_7$ factor gives (4, 6) points at the intersections of
the divisors associated with these factors (unless we are in the
special case where $-a \cdot b = b \cdot b = 0$ for these factors).
An interesting set of examples of this type with at least two $E_7$
factors were recently considered by Kimura in the F-theory context
\cite{Kimura}.  For example, the first case considered there has
$\alpha_4 = (t-x_1)^3 (t-x_2)^3 (t-x_3)^2 \prod_{i = 1}^{8} (u-y_i)$,
where $t, u$ are local coordinates on the base $\F_0 = \P^1 \times
\P^1$, giving a gauge group $(E_7^2 \times SO(7) \times
SU(2)^8)/\Z_2$.  The other models come from combining loci of the
SU(2) factors to form further $SO(7)$ or $E_7$ factors in the group.
While these models have (4, 6) points corresponding to SCFTs coupled
to the $SO(7)$ and $E_7$ factors, the analysis of Kimura suggests that
interpreting these loci as bi-charged matter may nonetheless give
solutions to the anomaly cancellation conditions.  The notion that
$({\bf 56}, {\bf 2})$ matter transforming under $E_7 \times SU(2)$ may
be realized in F-theory at apparent superconformal loci was also considered
previously in \cite{chl-exotic}.  As pointed out there, the existence
of a heterotic dual with such matter suggests that there is a sensible
interpretation of these constructions in the F-theory context.  It
would be interesting to investigate this kind of exotic matter
structure further.

\subsection{$\Z_3$ torsion: $SU(3), E_6, SU(6)$ (but not SU(9))}
\label{sec:z3}

A Weierstrass model with $\Z_3$ torsion takes the form
\cite{Silverman, Aspinwall-Morrison}
\begin{align}
f  &=-\alpha_1^4/48 +  \alpha_1 \alpha_3/2\nonumber\\
g & =  \alpha_1^6/864- \alpha_1^3 \alpha_3/24 + \alpha_3^2/4 \label{eq:z3}\\
\Delta & = \alpha_3^3 \tilde{\Delta}  \,.\nonumber
\end{align}
This implies at least an $\gsu(3)$ algebra.
This form of the Weierstrass model
can be understood geometrically by noting that an elliptic curve
only has a point $p$ satisfying $3 p = p + p + p = 0$ when   the point
$p$ is an inflection point that hits the tangent at that point three
times, in which case the Weierstrass model can be written in the form
$y^2 + a_1 y x + a_3 y = x^3$.

As for SU(2), we can easily prove that a model with algebra $\gsu(3)$
takes the form (\ref{eq:z3}) when there is no fundamental matter and $-a \cdot b >
  0$, as predicted by the massless charge sufficiency conjecture.
The general form of a Weierstrass model with $\gsu(3)$ algebra takes
the form \cite{mt-singularities}
\begin{align}
 f & = -\frac{1}{48} \phi_0^4 +   \frac{1}{2}
\phi_0 \psi_1 \sigma\nonumber + f_2 \sigma^2\\ g &
 =\frac{1}{864} \phi_0^6 -\frac{1}{24} \phi_0^3 \psi_1
\sigma + (\frac{1}{4} \psi_1^2 -\frac{1}{12} \phi_0^2 f_2)
\sigma^2
+ g_3 \sigma^3 \,.
\label{eq:su3}
\end{align}
The absence of fundamental matter dictates that $(-3a-b) \cdot b$ =
0.  The classes of $f_2, g_3$ are $[f_2] = -4a-2b,[g_3] = -6a-3b$, so
these are both ineffective and $f_2 = g_3 = 0$.
This gives exactly the form of (\ref{eq:z3}) with $\alpha_1 =\phi_0,
\alpha_3 = \psi_1 \sigma$, confirming that all models with no fundamental
matter and $-a \cdot b > 0$ have $\Z_3$ torsion, so that
the massless charge sufficiency conjecture holds for $\gsu(3)$

The simplest class of $\gsu(3)$ models of this form are those
 where the divisor $\sigma$
supporting the gauge factor has the maximum class $-3K_B$.
In this case $\alpha_3 = c \sigma$ with $c$ a constant.
In the case where  $\sigma$ is irreducible, this gives a 6D
F-theory model with gauge group $SU(3)/\Z_3$, and only adjoint matter
fields (with multiplicity $g = 1 + 3K_B^2$). For $T = 0$, this
reproduces the only anomaly-free theory with $\gsu(3)$ gauge algebra
and only adjoint matter.

As in the $\Z_2$ case, when we tune an SU(3) factor on a large enough
divisor $\sigma$ that there are no terms in $f, g$ of higher order in
$\sigma$, then we get the Weierstrass form (\ref{eq:z3}) with
$\alpha_3 = \tau \sigma$ and there is another gauge factor of SU(3) on
$\tau$ (multiple factors if $\tau$ is reducible), for a global gauge
group $(SU(3) \times SU(3))/\Z_3$.  As we found for the analogous
$\gsu(2)$ models, this construction also points to further classes of
anomaly-free models that have no known F-theory construction, such as
for example the $T = 0$ model with gauge group $SU(3)$, anomaly
coefficient $b = 8$, and 24 fields in the fundamental of SU(3) (along
with 21 adjoint fields), since e.g. attempting to tune such a
Weierstrass model automatically gives a second SU(3) factor, only
bifundamental and adjoint fields, and a global gauge group $(SU(3)
\times SU(3))\Z_3$.

One can again in this case construct various product models by taking
$\alpha_3$ to be reducible when $[\alpha_3] = -3K_B$.

We can also consider models with the group $E_6$, which has a center
$\Z_3$.  Like in the $E_7$ case above, we have
\begin{align}
  f & = \sigma^3 \tilde{f} \nonumber\\
  g & = \sigma^4 \tilde{g} \label{eq:e6}\\
  \Delta & =  27 \sigma^8 \tilde{g}^2 \nonumber \,.
% \label{eq:}
\end{align}
Fundamental matter appears at the locus where $\tilde{g}= 0$.  
We can  prove the massless charge sufficiency conjecture for
$E_6$ in a similar fashion to the preceding cases.  The absence of
fundamental matter is equivalent to the condition $(-3a-2b) \cdot b =
0$.  It follows that $\tilde{f} = 0$, and
there can be no higher-order terms in 
$\tilde{g}$ in an expansion in $\sigma$.
For Weierstrass models of the form (\ref{eq:e6}), with no terms in $g$
of order $\sigma^5$
the monodromy
condition for $E_6$ is that $\tilde{g}$ is a perfect square $\tilde{g}
= \chi^2$ (otherwise
we have a group $F_4$).  When this condition is satisfied,
(\ref{eq:e6}) takes the form (\ref{eq:z3}) with $\alpha_3 = \sigma^2
\chi$.
Thus, whenever we have a group $E_6$ with no fundamental matter and
$-a \cdot b > 0$ it follows that the Weierstrass model has the $\Z_3$
  torsion form (\ref{eq:z3}). This proves the massless charge
  sufficiency conjecture for the group $E_6$.

  In the simplest example of $E_6$ theories with $\Z_3$ torsion,
 $\tilde{g}$ is a constant, which occurs when
$\sigma$ is a section of $- 3 K_B/2$.  This is possible, for example, on
the base $\F_0 = \P^1 \times \P^1$, when $\sigma$ is a curve of
bidegree (3, 3).  In this case $\tilde{f}= 0$, and we match the
Weierstrass form (\ref{eq:z3}) with $\alpha_1 = 0, \alpha_3 =\sqrt{4
  \tilde{g}}\sigma^2$, precisely corresponding to the case where the
$E_6$ theory has only adjoint matter.

Finally, we consider one case with exotic matter: SU(6) with only
matter in the adjoint and 3-index antisymmetric representation.
These matter representations preserve the $\Z_3$ subgroup of the
discrete center $\Z_6$, so that the global form of the gauge group is
SU(6)$/\Z_3$.
From anomaly cancellation, this is possible when $-9 a \cdot b = 6  b
\cdot b$.  
The general form of Weierstrass model with SU(6) gauge group and
3-index antisymmetric representations was worked out in \cite{mt-singularities}
As
shown in \cite{Huang-Taylor-elliptic}, 
for such models with no antisymmetric matter,
a 
Tate-type tuning can be
realized through the general Weierstrass form
\begin{equation}
 y^2 + a_1 x y + \tilde{a}_3 \sigma^2 y = x^3 \,,
% \label{eq:}
\end{equation}
where fundamental matter occurs at the locus $\tilde{a}_3 = \sigma =
0$, so that the absence of fundamental matter occurs precisely when
$(-3a-2b) \cdot b = 0$.  This Tate tuning matches with the $\Z_3$
Weierstrass form (\ref{eq:z3}).  Here again (omitting the details,
which are similar to the other cases) the massless charge
sufficiency hypothesis holds and we have the appropriate Mordell-Weil
torsion group.
In the simplest cases, this corresponds to a situation in F-theory where
$[\sigma]= -3 K_B/2$, which can be realized for example on $\F_0$.

A similar construction would seem to possible for an SU(9) theory,
where we use $a_3 = \tilde{a}_3 \sigma^3$, with $[\sigma] = - K_B$.
This would seem to correspond to the structure needed for an SU(9)
theory with $SU(9)/\Z_3$ symmetry.  This could be imagined for example
to occur in a model
in which all the matter is in 3-index antisymmetric
representations.  As discussed in e.g.\ \cite{matter-transition},
however, this representation only seems to arise in the context of (4,
6) points associated with an SCFT.  This can be seen in the
Weierstrass model from the vanishing of $f, g, \Delta$ to orders 4, 6,
12 at the locus $a_1 = \sigma = 0$.  Nonetheless, this suggests that
these SCFT loci may have a natural interpretation as exotic SU(9)
matter in the F-theory context that preserves the $\Z_3$ discrete
central symmetry, like the $E_7 \times SU(2)$ matter discussed above
that preserves a $\Z_2$ symmetry. How the anomaly equations work out
in this case, however, is rather unclear,\footnote{In particular, if
  we take the base $\P^2$ and $\sigma$ to be a cubic, the generic
  matter content from anomaly cancellation would be $-9$ fundamental
  fields and $+9$ two-index antisymmetric fields.  These cannot be
  exchanged for only 3-index antisymmetric matter, however, as the
  3-index antisymmetric representation is anomaly equivalent to 5
  2-index antisymmetric fields and $-14$ fundamental fields.  Thanks
  to Patrick Jefferson for discussions on this issue.} and a further
analysis of this class of models is left for further investigation.

\subsection{Higher torsion groups}

\subsubsection{$\Z_4$}
Continuing in this vein we next consider the torsion group $\Z_4$.  In
this case we cannot simply realize this through a theory with gauge
group $SU(4)/\Z_4$ and only adjoint matter.  We can see this
since from the anomaly
conditions the number of two-index antisymmetric fields on a curve $C$
associated with anomaly coefficient $b$
is $-a \cdot b$, while the number of fundamental fields is $-8a \cdot
b-4b \cdot b$; these can only both vanish if $b \cdot b = -a \cdot b =
0$, which is an exceptional case where the condition $-a \cdot b > 0$
  does not hold.
The same kind of argument holds for all higher $SU(N)$, so we do not expect any
F-theory models with a gauge group of only $SU(N)/\Z_N$ for $N > 3$,
  except when $-a \cdot b = 0$, so the massless charge sufficiency
  condition holds trivially in this class of cases.

Similarly, we do not expect any models with torsion $\Z_4$ from a
gauge algebra $\gso(4n + 2)$.  The $\Z_4$ center is broken by
fundamental fields, with multiplicity $(-(2n-1)a - (2n-2)b) \cdot b$
and also by spinor fields, with multiplicity $(-2a -b) \cdot b$.
These multiplicities can only both vanish when $ -a \cdot b = b \cdot
b$, which is again one of the exceptional cases, so we again can have
no counterexamples to the massless charge sufficiency conjecture.

We can identify some models with torsion $\Z_4$ by analyzing the
Weierstrass model form with this torsion \cite{Aspinwall-Morrison}
\begin{equation}
 y^2 + a_1 x y + a_1 a_2 y = x^3 + a_2 x^2 \,,
% \label{eq:}
\end{equation}
which gives
\begin{align}
f & = - \frac{1}{48} a_1^4+ \frac{1}{3} a_1^2 a_2 - \frac{1}{3}
a_2^2\nonumber\\
g & = \frac{1}{864} (a_1^2 - 8 a_2) (a_1^4 -16 a_1^2 a_2 - 8 a_2^2)\\
\Delta & = - \frac{1}{16} a_1^2 a_2^4 (a_1^2 -16 a_2) \,.\nonumber
% \label{eq:}
\end{align}
From this we see that this torsion arises when the gauge algebra
contains at least $\gsu(2)\oplus\gsu(4)$.  In particular, when
$a_1, a_2$ are irreducible divisors in the classes $- K_B, - 2 K_B$ we get
a gauge group $(SU(2) \times SU(4))/\Z_4$.
Consideration of anomalies shows that the spectrum of an SU(2) gauge
group on a divisor of class $- K_B$ has $6 K_B \cdot K_B$ fields in the
fundamental representation, and the spectrum of an SU(4) gauge group
on a divisor of class $-2 K_B$ has $2 K_B \cdot K_B$ fields in the two-index
antisymmetric representation.  The curves $a_1, a_2$ intersect at $2 K_B
\cdot K_B$ points, so each of these points must support a
half-hypermultiplet in the $({\bf  2},{\bf 6})$ representation of
$\gsu(2)\oplus\gsu(4)$, which is invariant under $\Z_4$.  Indeed,
while we have not done an exhaustive analysis, it does not seem
possible to construct any other anomaly-consistent theory
that has a spectrum invariant under a
$\Z_4$ subgroup, other than by making the SU(2) and SU(4) loci $a_1$ and/or $a_2$
reducible.  Note in particular that trying to factor $a_2 = \alpha^2$
to get an SU(8) factor gives (4, 6) points at the locus $a_1 = \alpha
= 0$ unless $-a \cdot b = 0$, which is an exceptional case.
Similarly, factoring $a_2 = a_1 \beta$, which naively would give a
theory with gauge group $SU(4) \times SO(12)$ gives (4, 6) points at
the locus $a_1 = \beta = 0$ unless we are in an exceptional case.

\subsubsection{$\Z_2 \oplus \Z_2$}
\label{sec:z22}

In this case the Weierstrass model comes from a Tate form
\begin{equation}
 y^2 = x (x - b_2) (x - c_2),
% \label{eq:}
\end{equation}
giving
\begin{align}
  f & = \frac{1}{3} (b_2 c_2 - b_2^2 - c_2^2)\nonumber\\
  g & = - \frac{1}{27} (b_2+ c_2) (b_2-2 c_2) (2 b_2 - c_2) \label{eq:z2z2}\\
  \Delta & = - b_2^2 c_2^2 (b_2 - c_2)^2 \,. \nonumber
% \label{eq:}
\end{align}

The only single-factor groups that have a center $\Z_2 \oplus \Z_2$ are the
groups spin$(4n)$, with algebra $\gso(4n)$.  To keep this center
unbroken we must have no spinor matter, of multiplicity $(-2a-b) \cdot
b$, and no fundamental matter, of multiplicity $(-(2n-2)a -(2n -3) b)
\cdot b$.  This is only possible when $n = 2$, i.e. for $\gso(8)$.
From the point of view of Weierstrass models, the absence of spinor
matter, as discussed in \S\ref{sec:so}, is only possible with $-a
\cdot b > 0$ when we have an $\gso(8)$ algebra and a
  Weierstrass form (\ref{eq:so}) with $f_3 = g_4 = 0$, and $f_2 =
  cA^2, g_3 = c' A^3$.  This corresponds to a model with SO(8) gauge
  group and no fundamental or spinor matter.
This has the form (\ref{eq:z2z2}), where $c_2, b_2$ are proportional
to $A$,
and thus has the necessary $\Z_2 \oplus
\Z_2$ torsion, so the massless matter sufficiency conjecture is
satisfied for all single-factor groups with torsion $\Z_2 \oplus \Z_2$.

More generally, (\ref{eq:z2z2}) gives a gauge group of $SU(2)^3/(\Z_2
\oplus \Z_2)$, where the three SU(2) factors are tuned on divisors
$b_2, c_2, b_2 - c_2$ in the class $-2 K_B$.  The anomaly-free
spectrum consists of $4 K_B \cdot K_B$ half-hypermultiplet
bifundamental fields in each pair of SU(2) factors.  The $\Z_2$
actions act on pairs of SU(2) factors; note that the product of
e.g.\ the $\Z_2$ actions on factors 1+2 and 2+3 combine to give the
$\Z_2$ action on the 1+3 factors.

As in the other cases, we can get more complicated models by taking
$b_2, b_3$ in factored form.  
For example, setting $b_2 = \beta_1^2$ gives a theory
with $(SU(4) \times SU(2) \times SU(2))/(\Z_2 \oplus \Z_2)$ symmetry.

\subsubsection{Other discrete torsion groups}

For the groups $\Z_5, \Z_6, \Z_4 \oplus\Z_2,$ and $\Z_3 \oplus\Z_3$ 
there are no natural realizations of these torsion groups from
anomaly-free models with
single-factor gauge groups.  The only place these might be realized
from a single group factor is through $SU(5)/\Z_5$ or $SU(6)/\Z_6$,
but there are no anomaly-free models with these gauge groups and no
fundamental or two-index antisymmetric matter where $-a \cdot b > 0$.

Indeed, it turns out that there are no good Weierstrass models with
these Mordell-Weil torsion forms other than models with (4, 6) points
and models where gauge group factors are tuned on divisors that
violate the massless charge sufficiency condition $-a \cdot b > 0$,
  which we discuss further in Section \ref{sec:special-cases}.

As an example, for $\Z_5$, the Weierstrass model takes the form
\cite{Aspinwall-Morrison}
\begin{align}
f & = \frac{1}{6} a_1 b_1^3 - \frac{1}{48} a_1^{4} + \frac{1}{3} a_1^2
b_1^2 - \frac{1}{3} b_1^{4} - \frac{1}{6} a_1^3 b_1 \label{eq:z5}\\
g & = \frac{1}{864} (a_1^2 - 2 a_1 b_1+2 b_1^2) (a_1^{4} +14 a_1^3
b_1+26 a_1^2 b_1^2 -116 a_1 b_1^3 +76 b_1^{4}) \nonumber\\
\Delta & = \frac{1}{16} (a_1^2 +9 a_1 b_1-11 b_1^2) (a_1 - b_1)^{5}
b_1^{5} \,. \nonumber
% \label{eq:}
\end{align}
Here the parameters $a_1, b_1$ are both sections of the line bundle
${\cal O} (- K_B)$.  When $K_B \cdot K_B \neq 0$, then there are
points where both $a_1$ and $b_1$ vanish.  At such points, $f, g,
\Delta$ vanish to orders 4, 6, 12, corresponding to the appearance of
an SCFT in the 6D theory so that there is no simple interpretation in
terms of gauge groups and local matter fields.  We do not consider
such constructions here, although again the question of interpreting
SCFT loci as exotic matter fields begs further investigation.  Thus,
the only theories we consider that admit discrete $\Z_5$ Mordell-Weil
torsion are those where gauge group factors SU(5) or larger are tuned
on divisors satisfying $-a \cdot b = b \cdot b = 0$.  The same is true
for the other discrete torsion cases mentioned above: $\Z_6, \Z_4
\oplus\Z_2,\Z_3 \oplus\Z_3$.

\section{U(1) gauge factors}
\label{sec:abelian}

All the cases considered in the previous section involved nonabelian
gauge factors, associated with local Kodaira-type 
fiber singularities arising at codimension one loci in the base.  We
have given a fairly comprehensive treatment of such factors.  Abelian
U(1) factors, however, associated with the infinite $\Z^r$ part of the
Mordell-Weil group, are much less well understood.  The assertion of
massless charge sufficiency for the continuous part of the gauge
group states that the massless charges in the theory must span the
full charge lattice, even when such U(1) factors are included.
We conjecture that this holds for any U(1) factor with anomaly
coefficient $\tilde{b}$ satisfying $-a \cdot \tilde{b} > 0$.
Because the matter structure of U(1) theories is not as well
understood, it is harder to get a complete picture of this class of
theories.  
Here we consider only a single case: a situation where
there appears to be a U(1) factor with massless fields only of charge
2.  We show that in fact this case satisfies the massless charge
sufficiency hypothesis through the appearance of an additional
section that is the correct generator for the Mordell-Weil group.
A brief discussion of further cases involving U(1) gauge factors is
given in \S\ref{sec:abelian-more}

\subsection{U(1) and ``charge 2'' matter}

We thus now consider the case of the Morrison-Park model that appears to
have only charge 2 matter under the U(1) field (i.e., no matter of charge 1).
We first recall the general form of the Morrison-Park
\cite{Morrison-Park} model, which gives a generic
F-theory compactification with a U(1) gauge field and charges $q = 1,
2$, using the notation and conventions of \cite{mt-sections} (but
using $\hat{b}$ in place of $b$ to avoid confusion with the anomaly
coefficient $b$ for nonabelian groups)
\begin{equation}
 y^2 = x^3+ (- \frac{1}{3}e_2^2 +e_1e_3-e_0\hat{b}^2) x
 + (\frac{1}{3}e_1e_2e_3 +  \frac{2}{3}e_0e_2\hat{b}^2 -\frac{1}{4}e_1^2\hat{b}^2-
 e_0e_3^2  - \frac{2}{27}e_2^3) \,.
% \label{eq:}
\end{equation}
The parameters $e_i$ are sections of particular line bundles
parameterized by the single class $L$:
\begin{eqnarray}
\hspace*{0.1in}[e_0] & = & 2L\\
\hspace*{0.1in}[e_1] & = &  - K_B + L\\
\hspace*{0.1in}[e_2] & = &  -  2K_B\\
\hspace*{0.1in}[e_3] & = &  - 3K_B - L\\
\hspace*{0.1in}[\hat{b}] & = &  - 2K_B -L\,.
\end{eqnarray}
The anomaly coefficient $\tilde{b}$ of the resulting U(1) model
is associated with the class $2[e_3]$.
We are interested in particular in two special cases: In the first
special case, only massless matter with charge $q = 1$ appears.  This
occurs when $L = -2 K_B$, so that $\hat{b}$ is a constant, which we can
rescale so that $\hat{b} = 1$.  In the second
special case, where for clarity we denote the parameters by $\check{e}_i$,
only massless matter with charge $q = 2$ appears.  This
occurs when $L = K_B$, in which case $\check{e}_0 = 0$ must vanish identically
as $2 L = 2 K_B$ is not effective, and $\check{e}_1$ is a constant, which we can
take to be $\check{e}_1 = 1$.
In these two cases we have $\tilde{b} = -2a, \tilde{b} = -4a$.  As
long as $T < 9$, these give models satisfying $-a \cdot \tilde{b} > 0, \tilde{b} \cdot
\tilde{b} > 0$.

Our assertion is that the Morrison-Park models in these two special
cases are precisely equivalent.  That is, the $q = 2$ model with
\begin{align}
  f & = \check{e}_3 - \check{e}_2^2/3\label{eq:q2}\\
  g & =- ( - \frac{2}{27} \check{e}_2^3 + \frac{1}{3} \check{e}_2 \check{e}_3 -
  \frac{1}{4} \check{b}^2)\nonumber
\end{align}
is equivalent to the $q = 1$ model with
\begin{align}
  f & = - \frac{1}{3}e_2^2 +e_1e_3-e_0\label{eq:q1}\\
  g & = - (\frac{1}{3}e_1e_2e_3 + \frac{2}{3}e_0e_2 -\frac{1}{4}e_1^2-
e_0e_3^2  - \frac{2}{27}e_2^3) \,.
\nonumber
\end{align}
%\wati{I think these signs are all correct now but should confirm.}
This equivalence occurs because the natural section for the $q = 2$
special case of Morrison-Park is actually not the generating section
of the Mordell-Weil group;
as noted in the final appendix of \cite{Morrison-Park}, this model
contains an additional section corresponding to the generator of the
U(1), which matches the natural section in the $q = 1$ special
case.\footnote{We would also like to thank Remke Kloosterman and
  Nikhil Raghuram for discussions regarding this additional section.}
To show this equivalence explicitly, we identify the relationship
between the parameters in the two models to be
\begin{align}
  \check{b} & = - e_1+2 e_2 e_3-2 e_3^3\nonumber\\
  \check{e}_2 & = -2 e_2+3 e_3^2\label{eq:equivalence}\\
  \check{e}_3 &= - e_0+ e_2^2+ e_1 e_3-4 e_2 e_3^2 + 3 e_3^4 \,.
\nonumber
\end{align}
Explicit use of these relations shows through a short calculation that
the models are indeed equivalent as stated.  Furthermore, by taking
$e_3$ to be an arbitrary section of ${\cal O} (- K_B)$, we can solve the
equations for $e_2, e_1, e_0$ successively, so that for every $q = 2$
type model there is a family of $q = 1$ type models associated with
the allowed values of $e_3$, representing an over-parameterization of
the associated Weierstrass models.

This shows that the models that appear to have only $q = 2$ charges
under the U(1) gauge symmetry are actually equivalent to models with
the fundamental charge $q = 1$.  Thus, the ``$q = 2$'' type models
actually have an additional section $\sigma$, as expected
through
Poincar\'{e} duality and the massless charge sufficiency
hypothesis; this section satisfies $\sigma \cdot C = 1$ for the
curves that appear to have charge 2 in the  usual Morrison-Park
formulation.

While this explicit demonstration of equivalence is mathematically
efficient, it does not explain much about the geometry.  We make some
additional observations that help to clarify this story.

First, we note that the $q = 1$ case, where $\hat{b}$ is a constant that can
be taken to be unity, is the only case where we can shift the section
identified by Morrison-Park
\begin{equation}
[x_1, y_1, z_1] = [e_3^2 - \frac{2}{3} \hat{b}^2e_2, -e_3^3 + \hat{b}^2e_2e_3
-  \frac{1}{2} \hat{b}^4e_1, \hat{b}]
% \label{eq:}
\end{equation}
to the origin without involving rational functions with a nontrivial
denominator.  Performing the shift
\begin{equation}
 x \rightarrow x + x_1,  \; \;y \rightarrow y + y_1
% \label{eq:}
\end{equation}
the $q = 1$ Weierstrass model takes the form
\begin{equation}
 y^2 + a_3 y = x^3 + a_2 x^2 + a_4 x \,,
\label{eq:shifted}
\end{equation}
where the section is now located at $x = y = 0$, and we
identify (setting $\hat{b} = 1$)
\begin{align}
  a_2  & = 3 x_1 \nonumber \\
  a_3 & = 2 y_1 \label{eq:ax}\\
  a_4 & = - e_0+ e_2^2+ e_1 e_3-4 e_2 e_3^2 + 3 e_3^4 \,. 
\nonumber
\end{align}
If we now complete the square and cube in the Weierstrass model
(\ref{eq:shifted}), we find that the resulting Weierstrass normal form
has precisely the form of (\ref{eq:q2}), with the identifications
\begin{align}
  \check{b} & = a_3 \nonumber \\
  \check{e}_2 & = a_2\label{eq:ba}\\
  \check{e}_3 & = a_4 \,.\nonumber
\end{align}
On the other hand, completing the square in this way must return us to
the original $q = 1$ Weierstrass model.  Combining (\ref{eq:ba}) and
(\ref{eq:ax}), we reproduce the relations of (\ref{eq:equivalence}),
giving a simple way of understanding the formulae appearing in this
relationship.

To confirm this picture, we can check explicitly
that the natural section of the $q = 2$ Morrison-Park model
corresponds under the elliptic curve addition law to twice the natural
section of the $q = 1$ model.  In terms of the parameterization
(\ref{eq:shifted}), where the section $ s = [x_1,  y_1, 1]$ has been
shifted to the origin we can compute $s + s$ using the elliptic curve
law.  The tangent to (\ref{eq:shifted}) at the point $s$ is given by
the line $y = a_4 x/a_3$.  This line intersects the cubic at the third
point $\tilde{r}= (\tilde{x}, a_4 \tilde{x}/a_3)$, where $\tilde{x}= -
a_2+ a_4^2/a_3^2$.  This represents the point $- 2 s$ on the elliptic
curve.  To find $2 s$, we identify the other point on the cubic that
intersects a vertical line passing through this point, which is
\begin{equation}
 2 s \sim (- a_2+ a_4^2/a_3^2, - a_3+ a_2 a_4/a_3 - a_4^3/a_3^3)
% \label{eq:}
\end{equation}
Shifting $x, y$ to go back to the original Weierstrass normal form,
and scaling by appropriate powers of $a_3$, we see that this precisely
reproduces the expected Morrison-Park section
\begin{equation}
  2 s \rightarrow [- \frac{2}{3} a_2 a_3^2 + a_4^2,
    - a_3^4/2+ a_2 a_4 a_3^2 - a_4^3, a_3] \,.
% \label{eq:}
\end{equation}
So we find that indeed the natural section of the ``$q = 2$''
Morrison-Park model is twice the generating section identified from
the ``$q = 1$'' form of the same model, confirming the massless charge
sufficiency hypothesis in this situation.

\subsection{Other aspects of U(1) factors}
\label{sec:abelian-more}

We have considered here only a single example of a theory with a U(1)
factor that appears naively to violate the massless charge sufficiency
conjecture.  There are many other similar models that satisfy anomaly
cancellation conditions \cite{TT-infinite}; for example, any model
with 54 matter fields with each of the U(1) charges $aq, ap, a (q +
p)$ satisfies anomaly cancellation, for arbitrary integers $a > 1, p, q$,
but violates the massless charge sufficiency conjecture.  Few such
models have known F-theory constructions, however, and F-theory in
principle can only realize a finite subset of this infinite class of models.
In particular, explicit Weierstrass models are only known for U(1)
charges up to $q = 3, 4$ \cite{Klevers-3, Raghuram-34}, though general
arguments suggest that charges as large as $q = 6$
\cite{Cianci-mv, Collinucci-high} or even $q = 21$ \cite{Raghuram-WT-large}
should be possible in 6D F-theory models.
It was shown in \cite{Raghuram:2020vxm} that the above-mentioned
infinite family of U(1) models would appear to violate the automatic enhancement
conjecture, and indeed lead to a larger gauge group in the simplest
cases where F-theory realizations are possible, but there are other more complicated infinite families where
this story is even less clear.
It would be interesting to
explore further if and how massless charge sufficiency is realized
for models that apparently have matter content where all charges under
a given U(1) factor are multiples of an integer $a > 2$.

Another interesting class of models that we have not addressed in the
preceding sections involve
situations where the gauge group has nonabelian and abelian factors,
and may have a discrete quotient, such as $U(N)=(SU(N) \times U(1))/\Z_N$.
Some analysis of the charges in such situations were considered in
\cite{Lawrie-sw, Grimm-kk, Cvetic-Lin}.  Models of this type where the
gauge group contains $SU(3) \times SU(2) \times U(1)$ factors, 
with the possibility of a discrete quotient by $\Z_6$, have recently
been considered in \cite{Lin-Weigand, Cvetic-kpor, Lin-Weigand-2,
  Cvetic-llo, Cvetic-hllt, TT-sm, RTT}.  In this situation in particular, for
example, however, as discussed in more detail in the appendix, the
fundamental group of the gauge group $(SU(3) \times SU(2) \times
U(1))/\Z_6$ is just $\Z$, and there is no Mordell-Weil torsion.  The
generic matter representations under this group, which include the
charges of the MSSM \cite{TT-generic}, generate the full charge
lattice of the quotient group, and the massless charge sufficiency
hypothesis is satisfied for the models that have been considered of
this type.

\section{Exceptions to massless charge sufficiency}
\label{sec:special-cases}

As discussed in Section \ref{sec:hypotheses}, there are two distinct
classes of exceptions to the condition $- a \cdot b > 0$ that we
conjecture leads to massless charge sufficiency.  The first class
corresponds to nonabelian factors where $- a \cdot b <0$, and the
second class corresponds to abelian or nonabelian factors where $- a
\cdot b = b \cdot b = 0$.  We consider these two cases of exceptions
in turn.

\subsection{Non-Higgsable nonabelian factors: exceptional cases with
  $-a \cdot b < 0$}
\label{sec:exceptional-non-Higgsable}

The exceptional cases with $- a \cdot b <0$ correspond to F-theory
models  where the base $B$ contains rational curves with $- K_B \cdot C <0$.
Multiples of the anticanonical class $- n K_B$ must contain
one or more multiples of the  rigid curve $C$, which has
self-intersection $C \cdot C = - K_B \cdot C - 2$, as a component.  It
follows that the Weierstrass coefficients $f, g$ must vanish over $C$
to sufficiently high degrees to force at least a type IV vanishing
associated with an $\gsu(3)$ algebra.  The configurations of curves of
this type that are allowed in F-theory models without (4, 6) loci are
known as ``non-Higgsable clusters,'' since the associated gauge groups
cannot be broken by giving a Higgs VEV to any of the matter fields
without breaking supersymmetry, and were classified in
\cite{clusters}.

Many of the non-Higgsable clusters give single gauge factors with no
matter.  In particular, curves of self-intersection $- 3, - 4, -6, -8,
-12$ give rise to the gauge algebras $\gsu(3),\gso(8), \age_6, \age_7$,
and $\age_8$.  While there is no matter charged under any of these
gauge fields, in each of these cases the constraints on the
Weierstrass model are completely local and in general there is no
Mordell-Weil torsion.  
As a simple example of this consider the SU(3) non-Higgsable gauge
factor that arises on the $- 3$ curve $S$ in the Hirzebruch surface
$\F_3$.  Standard toric methodology (see, e.g., \cite{toric}) gives an
explicit description of $f, g$ in terms of monomials associated with
lattice points in the 4th and 6th polar polytopes to the polytope
associated with the toric base $\F_3$.  Taking $\sigma$ to be a
coordinate that vanishes on $S$, we have
\begin{align}
  f & = f_{2, 2} \sigma^2 + f_{3, 5} \sigma^3 + \cdots\\
  g & = g_{2, 0} \sigma^2 + g_{3, 3} \sigma^3 + \cdots
% \label{eq:}
\end{align}
where the second index in the subscript denotes the degree of the
associated coefficient considered (locally) as a function of the coordinate on
the rational curve $S$.
The coefficients of each of the monomials in this expansion can be
chosen independently, as $f, g$ are general sections of $- 4 K_B, - 6
K_B$ respectively.
It is easy to compute that the numbers of independent degrees of
freedom in $f, g$ are $84, 176$ respectively (this is closely related to the
total number of uncharged hypermultiplets in the theory, 252).
On the other hand, for a Weierstrass model with $\Z_3$ torsion, which
must have the form (\ref{eq:z3}), the parameters $\alpha_1, \alpha_3$
are sections of $- K_B, - 3 K_B$ and have only $9, 51$ degrees of
freedom respectively.  Thus, the generic Weierstrass model over $\F_3$
has an $\gsu(3)$ gauge algebra over $S$ but does not have Mordell-Weil
torsion, so the gauge group is indeed SU(3).  Since there is no matter
this represents a case in which the massless charge sufficiency
condition does not hold.  Similarly, since the other single
non-Higgsable gauge group factors without matter result from purely
local effects in the geometry they do not generically correspond to
Weierstrass models with Mordell-Weil torsion, which is a global
effect.  The other single non-Higgsable gauge factors are therefore
spin$(8), E_6, E_7,$ and $E_8$.  All of these except for the $E_8$
case represent exceptions to the massless charge sufficiency condition.

Over a $- 5$ curve there is a gauge factor $F_4$.  This does not have
a nontrivial discrete center so there is no possibility of
Mordell-Weil torsion.  Over a $- 7$ curve there is a $E_7$ gauge group
with a single multiplet of matter in the fundamental (56).  Because
the matter breaks the central $\Z_2$, there is no possibility for
Mordell-Weil torsion.

There are also three non-Higgsable clusters containing a combination
of $-3$ and $- 2$ curves.  A cluster containing one each of these
curves, intersecting at a point, supports a gauge group $G_2 \times
SU(2)$, with a half-hypermultiplet of matter in the (7, 2)
representation.  This breaks the $\Z_2$ central symmetry of the SU(2)
factor so there is no possibility of torsion.  The same situation
holds for a chain of three curves of self-intersections $- 3, - 2, -
2$.  In the final case, however, we have a chain of three curves of
self-intersections $- 2, - 3, - 2$, supporting a gauge group $SU(2)
\times SO(7) \times SU(2)$.  There is a matter field jointly charged
under the SO(7) factor and each of the SU(2) factors, in the (8, 2)
representation.  This matter breaks two of the three combinations of $\Z_2$ factors in
the center.  There is, however, a third $\Z_2$ factor in the center,
given by the product of all three $\Z_2$ factors, which acts trivially
on all the matter in this cluster.  Nonetheless, like the single gauge
factors this is a local effect and there is no Mordell-Weil torsion in
general, so this cluster represents another exception to massless
charge sufficiency.

Finally, we point out that the Spin(8) on a $- 4$ curve can be enhanced
to a larger Spin$(N)$ group that has fundamental matter but no spinor
matter.  In this case there can still be a discrete subgroup of the
gauge factor that acts trivially on all the matter, but the gauge
group is still Spin$(N)$; this gives further exceptions to massless
charge sufficiency where $- a \cdot b <0$.

This essentially exhausts all the possible exceptions in the first class.

\subsection{Geometry of exceptional cases with $-a \cdot b = b \cdot b
  = 0$}

In this section we describe in further detail the second class of
exceptions to the massless charge sufficiency hypothesis for 6D
F-theory models.  These exceptions arise in cases where we have a
gauge factor associated with an irreducible divisor $C$
such that
\begin{equation}
K_B \cdot C = C
\cdot C = 0\,.
\label{eq:exception-condition}
\end{equation}
Physically, in these cases the matter content matches
with what we would expect in a theory with twice as much
supersymmetry, where the vector supermultiplet contains a single
adjoint field.  For a nonabelian gauge factor this means that the matter
content in the 6D ${\cal N} = 1$ theory contains a single adjoint
hypermultiplet, and for an abelian U(1) factor it means that there is
no matter charged under the U(1).  In both cases, there is no one-loop
anomaly associated with the gauge factor, and no Green-Schwarz
mechanism occurs, in accord with the vanishing of the terms $K_B \cdot
C, C \cdot C$ that appear on the left-hand side of the anomaly equations.

Geometrically, the significance of the conditions
(\ref{eq:exception-condition}) is that the curve $C$ is a genus one
curve (i.e.\ a two-dimensional torus) with trivial normal bundle.
Locally, therefore, the geometry is like that of a flat 7-brane in 10D
flat space, so it makes sense that the matter content matches with
that expected of the vector supermultiplet with enhanced
supersymmetry. The existence of such a genus one curve with trivial
normal bundle means that, at least to first order locally, the base $B$ is itself
elliptically fibered.

The simplest base $B$ that contains such a curve $C$ is the del Pezzo
surface $dP_9$, also known as a rational elliptic surface.  This
surface can be thought of as an elliptic fibration over the 1D base
$\P^1$, where instead of the 24 vanishing points of the discriminant
needed to form a K3 surface, the discriminant vanishes only to order
12, and $f, g$ are sections of ${\cal O} (-2K_{\P^1})$ and  ${\cal O}
(-3K_{\P^1})$ respectively.  Other bases can also contain a curve
satisfying
(\ref{eq:exception-condition}), e.g.\ by blowing up a point on the base
well away from $C$ (which we note is only possible when we have a
degenerate surface containing curves of self-intersection $-3$ or
below, which support non-Higgsable  nonabelian gauge factors).

\subsection{Non-Higgsable U(1) factors}

The generic elliptic fibration over a smooth rational elliptic surface
gives a Calabi-Yau threefold with Hodge numbers (19, 19) known as a
Schoen manifold.\footnote{The generic Schoen manifold has an {\it
    infinite} number of distinct elliptic fibrations, see
  e.g. \cite{aggl} for a recent analysis in the context of F-theory.}  
% WT added reference
More generally, Schoen constructed a class of
Calabi-Yau threefolds as fiber products of rational elliptic surfaces;
these rational elliptic surfaces can include singular fiber
configurations, giving to a range of threefold constructions.  The
F-theory physics associated with such constructions was analyzed in
some detail in \cite{MPT}.  The most important feature of the Schoen
manifold is the presence of extra sections of the threefold elliptic
fibration at generic points in the moduli space.  The generic Schoen
manifold has 9 independent sections, leading to a fibration with
Mordell-Weil rank 8.  Physically, these sections give rise in F-theory
to U(1) gauge factors that are (supersymmetrically)
non-Higgsable; there are no massless
fields charged under these U(1) factors, and the U(1) factors are
present everywhere in the smooth moduli space.
These kinds of non-Higgsable U(1) factors were analyzed in
\cite{Martini-WT, Wang-u1, MPT}.
When the fiber product
includes rational elliptic surfaces with singular fibers, the sections
can combine into Kodaira singularities of the threefold associated
with nonabelian gauge factors; in this subsection we focus on the generic
non-Higgsable U(1) factors and look at the associated nonabelian
factors in the next subsection.

From the complete absence of massless fields charged under such a U(1)
factor, it is clear that they violate the massless charge sufficiency
conjecture.  It is interesting that such fields can only arise in the
``higher supersymmetry'' context where (\ref{eq:exception-condition})
is satisfied.

One might wonder how the absence of massless matter charged under the
U(1) factor is reconciled with charge completeness, which we have
proven using Poincar\'{e} duality. To demonstrate how this is possible
we carry out the Poincar\'{e} duality analysis of divisors and curves
a little further in this case.  As described in
\S\ref{sec:completeness-proof}, we begin by projecting out the
divisors $\sigma_\infty, \pi^{*}C$ and the curves $F, \hat{C}$.  In
the case of the generic Schoen manifold $X$ there are 8 further sections
$\sigma_i, i = 1, \ldots, 8$, where the Mordell-Weil group $\Z^{8}$ is
generated by $\sigma_i - \sigma_\infty$.
Subtracting off $\sigma_\infty$ from these sections ensures that we
have divisors that are orthogonal to the curve associated with the
generic fiber $F$.
We must further orthogonalize these divisors
by subtracting appropriate multiples of the divisors $\pi^{*}C_\alpha$
so that they are orthogonal to the curves $\hat{C}_\beta$.  This can
be done by taking the divisors
\begin{equation}
 D_i = \sigma_i - \sigma_\infty - \pi^{*}K \,.
% \label{eq:}
\end{equation}
From Poincar\'{e} duality, we expect that there are fibral curves
on $X$, even after the above projections,  that have nonzero inner
product  with $D_i$,
associated
with matter fields in the 6D theory that are charged under each of the
U(1) factors.  We can identify some such curves by starting with
$\pi^{*}C_\alpha \cdot \sigma_i$ and orthogonalizing with respect to
the divisors $\sigma_\infty, \pi^{*}C_\beta$.  The resulting curves
are of the form $\pi^{*}C\cdot (\sigma_i - \sigma_\infty) + (K_B \cdot C)
F$.  Note that these are fibral curves in homology but are expressed
as linear combinations of holomorphic curves that are horizontal in
the elliptic fibration, and do not have vanishing volume in the
F-theory limit, so that these correspond to massive  particles in the
6D F-theory model that are charged under the U(1) factor associated
with divisor $D_i$.  In this way, we can satisfy charge completeness
but not massless charge sufficiency for these non-Higgsable U(1)
factors that satisfy (\ref{eq:exception-condition}).  Note, however,
that the inner product between such a curve and $D_i$ is even and
given by $2 K_B \cdot C$.  So we have not given a complete set of
divisors and dual curves.  Other curves may for example be associated
with the lift of a curve in the base to a section that is only defined
over that curve and not globally over the base; it is also possible
that the curves we have identified here are even in integer homology,
and thus there exist curves associated with half of these classes.  
% WT added this possibility
We do not attempt a
complete analysis of the set of allowed curves here.

\subsection{Nonabelian factors with a single adjoint matter field}

The non-Higgsable U(1) factors described in the previous subsection
are closely related to nonabelian factors that satisfy
(\ref{eq:exception-condition}).  In particular, given any nonabelian
factor with only a single adjoint field, we can Higgs the nonabelian
factor to a product of U(1)s under which there is no charged matter.
For example, given an SU(2) factor with a single adjoint we can turn
on a diagonal expectation value for the adjoint reducing to a single
U(1) under which no matter is charged.  Geometrically, such nonabelian
gauge factors arise in the Schoen construction when one of the factors
in the fiber product of rational elliptic surfaces has degenerate
fibers.  The set of possible degenerate fibers for a rational elliptic
surface is known and was tabulated in \cite{Persson, Miranda}.
We can also identify
the gauge algebras that can appear on a divisor satisfying
(\ref{eq:exception-condition}) from the geometry of rational elliptic surfaces.  The intersection form on $dP_9$
is $U \oplus E_8$.
An elliptic Calabi-Yau threefold with base $dP_9$ is generally a fiber
product of rational elliptic surfaces.  The elliptic fiber of the
surface with which we take the fiber product removes the $U$ factor from the
intersection form.
This shows that the only gauge
algebras we can realize are those with a lattice that is a sublattice
of the $E_8$ lattice.  Furthermore, this gives information about the
kinds of discrete Mordell-Weil torsion that can be realized even in
these exceptional cases.

As an example, consider what happens in the $\Z_2$ torsion case
described by the Weierstrass model with (\ref{eq:z2}) when the base is
a generic rational elliptic surface $dP_9$.  In this case, $[\alpha_4]
= -4 K_B$, and $K_B$ is the class of the fiber with $K_B \cdot K_B = 0$.  This
means that $\alpha_4$ is reducible and factorizes as $\alpha_4 =
\beta_1 \gamma_1 \delta_1 \epsilon_1$.  The gauge group then becomes
$SU(2)^{4}/\Z_2$.  The presence of the $\Z_2$ quotient can be
understood from the fact that if $x, y, z, t$ are orthogonal elements
of the $E_8$ lattice each with self-intersection $- 2$, then $(x + y +
z + t)/2$ is also an element of the lattice with the same
self-intersection since $E_8$ is unimodular.

The upshot of this is that for the exceptional cases satisfying
(\ref{eq:exception-condition}), the presence of Mordell-Weil torsion
is determined by different considerations to the general 6D ${\cal N}
= (1, 0)$ gauge factor where we believe massless charge sufficiency
holds.  In some cases there is no Mordell-Weil torsion, though in
others there is.  Another example of a case with Mordell-Weil torsion
with a rational elliptic surface base is the case of
$\Z_5$ torsion (\ref{eq:z5}).  In this case we see that one can have
two SU(5) factors each on a divisor of class $- K_B$, with global
gauge group $(SU(5) \times SU(5))/\Z_5$.

%------------------------------------------------------------
\section{Conclusions}
\label{sec:conclusions}

In this paper we have used Poincar\'{e} duality on an elliptic
Calabi-Yau threefold to prove that the charge completeness hypothesis
for the connected component of the gauge group of every 6D ${\cal N} =
(1, 0)$ supergravity theory that is realized in F-theory is equivalent
to the standard understanding \cite{Aspinwall-Morrison,LieF,torsion-MW}
of how the global form of the gauge
group is encoded in F-theory geometry.  The same result holds for
every 4D ${\cal N} = 1$ F-theory model, for all charges under the
connected component of the gauge group that arises from 7-branes and
from sections of the  F-theory elliptic fibration.

We have conjectured a stronger condition for 6D supergravity theories,
that the massless charged states generate the full charge lattice of
the theory whenever the anomaly coefficients satisfy $-a \cdot b > 0$,
which holds in all but a small set of exceptional cases.  We have
checked this for essentially all single-factor nonabelian gauge
groups, and analyzed a large number of examples that illustrate the
connection between the fundamental group of the gauge group of the
supergravity theory and Mordell-Weil torsion in the F-theory picture.
We have found that the only possible torsion subgroups that can arise
without codimension two (4, 6) loci associated with superconformal
sectors are $\Z_2$, $\Z_3$, $\Z_4,$ and $\Z_2 \oplus \Z_2$, except in the
exceptional cases that violate massless charge sufficiency.  We have
given one example of an abelian U(1) theory that satisfies the
massless charge sufficiency condition in a nontrivial way, but a more
thorough analysis of more complicated U(1) theories with higher
apparent charges is left as an open question.

An interesting feature we have identified in several of the models
with fixed Mordell-Weil torsion groups is the appearance of SCFTs at
points where anomaly cancellation and the discrete torsion structure
suggest specific types of massless matter.  In one case, corresponding
to a situation previously analyzed in \cite{chl-exotic, Kimura}, we
find loci associated with $(E_7 \times SU(2))/\Z_2$ matter in the
({\bf 56}, {\bf 2}) representation.  These structures suggest that
there may be a general way of understanding certain exotic matter
structures in terms of local SCFTs in the F-theory context, as also
suggested in \cite{chl-exotic}. This would be interesting to
investigate further.

It would be nice to find some way to prove the charge completeness and
massless charge sufficiency hypotheses for
general 6D supergravity theories, extending the corresponding results of
\cite{Harlow-Ooguri} to flat space theories in 6D.

Even in F-theory, we do not have a proof of charge completeness for
theories with discrete gauge groups.  Poincar\'{e} duality does not
seem to be adequate to generalize the proof to such situations.  We
leave this interesting question, and a further analysis of 4D
theories, for further work.

While we have analyzed a broad range of examples, the appearance of
Mordell-Weil torsion in precisely those cases where the massless
charge sufficiency conjecture predicts it should arise seems
somewhat surprising and mysterious.  It seems there should be some
more general principle that guarantees that the global gauge group
indeed matches the massless matter when $-a \cdot b > 0$.  It would be
good to find a more fundamental understanding of this result, and to
better understand the exceptional cases that violate the massless
charge sufficiency condition when $-a \cdot b \leq 0$.
Even lacking a proof of the massless charge sufficiency hypothesis
for 6D theories, it would be interesting to find arguments based on
basic principles of quantum gravity such as using black holes, along
the lines of arguments in e.g. \cite{Banks-Seiberg}.

\vspace{.5in}

{\bf Acknowledgements}: We would like to thank 
Paul Aspinwall,
Daniel Harlow,
Patrick Jefferson, 
Remke Kloosterman, 
Nikhil Raghuram, 
Andrew Turner,
and
Timo Weigand
for helpful discussions.  WT would like to thank the Kavli Institute for
Theoretical Physics for hospitality on several visits during the
progress of this project.  We would also like to thank the Aspen
Center for Physics for hospitality during the early and final stages of this
project, and the Witwatersrand (Wits) rural facility and the MIT
International Science and Technology Initiatives (MISTI)
MIT-Africa-Imperial College seed fund program for hospitality and
support during the latter stages of this project.  The work of DRM was
supported in part by National Science Foundation grant PHY-1620842
(USA).  The work of WT was supported by the DOE under contract
\#DE-SC00012567.  This research was also supported in part by NSF
grant No.\ PHY-1748958.

\appendix

\section{The fundamental group of the gauge group}

In this appendix, we will review some aspects of the topology of gauge groups,
and describe how that topology is  
connected  to the representation theory of the gauge group,
in particular explaining why the fundamental group of the gauge group
gives a reasonable measurement of the representation theory.
Our primary references will be \cite{MR1997306} and \cite{MR1410059}. 

The gauge group $G$ of a physical theory is a compact Lie group 
%$G$
which in general might not be connected.  
Let $G^0$ be the connected component
of $G$ containing the origin.  Then $G^0$ is a normal subgoup of
$G$, and there is a short exact sequence of groups
$$ 1 \to G^0 \to G \to G/G^0 \to 1 $$
where $1$ denotes the trival group, and
where $G/G^0=\pi_0(G)$ is a discrete group, the {\em group of components
of $G$}.  As in the body of this paper, we ignore questions about
this discrete group, and only consider the connected subgroup
$G^0$.

It is known \cite[p.~70]{MR1997306} that a compact Lie group is connected if 
and only if it is path-connected.  So $G^0$ is also path-connected. 

\begin{theorem}[{\cite[Thm V(8.1)]{MR1410059}}]  A compact connected Lie group
$G^0$ possesses a finite (unramified) cover which is isomorphic to the
direct product of a 
compact simply connected Lie group $G_0$ and a (real) torus\footnote{In this
appendix, we follow the conventions of the math literature and refer to this torus 
as $\torus$; in the body of the paper, we have often remained closer to the physics
literature and called the torus $U(1)^r$.} $\torus\cong U(1)^r$.
\end{theorem}

Ref.~\cite{MR1410059} goes on to note that the covering map is given by the quotient by
a finite central subgroup $\Xi$, namely:
\begin{equation}\label{quotient}
 G^0 = (G_0 \times \torus)/\Xi
\end{equation}
where $\Xi$ is a finite subgroup of the center $Z_0\times \torus$
of $G_0\times \torus$.  Moreover, $G_0$
is a product of simply connected simple compact groups $G_i$
\begin{equation}
G_0 = \prod_{i=1}^k G_i
\end{equation}
and the center of $G_0\times \torus$ is $(\prod_{i=1}^k Z_i) \times \torus$, where
$Z_i$ is the (finite) center of $G_i$.  These centers are closely related
to the representation theory.

Before continuing with the general theory, we pause to discuss two familiar 
examples. If $G^0=U(n)$, the finite cover in question is obtained by extracting
an $n^{\text{th}}$ root of the determinant.  Then $(\det A)^{-1/n} A\in
SU(n)$ for $A\in U(n)$ and we can describe an element of $U(1)\times SU(n)$
as $((\det A)^{1/n}, (\det A)^{-1/n} A)$.  There is an order $n$ ambiguity
here (the choice of $n^{\text{th}}$ root) and so 
$U(n) \cong (U(1)\times SU(n))/(\mathbb{Z}/n\mathbb{Z})$, where the action of 
the central subgroup $\Xi\cong\mathbb{Z}/n\mathbb{Z}$ is generated by
$(e^{-2\pi i/n},\operatorname{diag}(e^{2\pi i/n}, \dots, e^{2\pi i/n})).$  
A representation of $U(1)\times SU(n)$  descends to a representation
of $U(n)$ if and only if
the group $\Xi$  acts trivially within the given representation.

As a second example, let $G^0$ be the gauge group of the standard 
model.\footnote{The global structure of this group has various subtleties,
and the corresponding representation theory is described in great
detail in
\cite{Baez:2009dj}.}
We take $G^0$ to be the subgroup of $SU(5)$ consisting of matrices in block
$2\times2$ and $3\times3$ form.  (The fact that the matter representation
of the standard model is compatible with this description is at the 
mathematical heart of the Georgi--Glashow grand unified model
\cite{Georgi:1974sy}.)  That is, a gauge group element
 consists of a unitary $2\times2$ matrix $A$ and a unitary $3\times 3$
matrix $B$ such that the $5\times 5$ matrix 
\[ \begin{bmatrix} A & 0 \\ 0 & B \end{bmatrix}\]
 has
determinant $1$, i.e., $(\det A)(\det B)=1$.
To describe this group in terms of $SU(2)$ and $SU(3)$, we need a square root of
$\det A$ and a cube root of $\det B$; since these determinants
were inverses of each other,
we need a sixth root of the common quantity.

That is, we can describe our group $G^0$ as a finite quotient of 
$U(1)\times SU(2) \times SU(3)$, where the map to $SU(5)$ is
given by 
%\[ (\lambda,\alpha,\beta) \mapsto \operatorname{block} (\lambda^3\alpha,
%\lambda^{-2}\beta). \]
\[ (\lambda,\alpha,\beta) \mapsto \begin{bmatrix}\lambda^3\alpha & 0 \\
0 & \lambda^{-2}\beta \end{bmatrix}. \]
It is easy to see that the kernel of this map is 
$\Xi\cong\mathbb{Z}/6\mathbb{Z}$ generated by 
\[\left(e^{-2\pi i/6},\operatorname{diag}(e^{6\pi i/6},e^{6\pi i/6}),
\operatorname{diag}(e^{-4\pi i/6},e^{-4\pi i/6},e^{-4\pi i/6})\right).\]
Thus, $G^0 \cong (U(1)\times SU(2)\times SU(3))/(\mathbb{Z}/6\mathbb{Z})$. 
The key fact here is that $\Xi$ acts trivially on all representations
ocurring in the standard model, and it is the largest subgroup of
$U(1)\times SU(2) \times SU(3)$ to do so.
 
Returning to the general case: 
the connected part $G^0$ of the gauge group will act on the various fields
of the physical theory\footnote{For simplicity, we take all fields to
be complex-valued, so that tori act through multiplication by some phase.
In the real case, a phase rotation in the complex $z$-plane can be replaced
by a circle action in the real $(x,y)$-plane.}, 
and the charges under this group are determined
by the charges under a maximal (compact) torus $H\subset G^0$.  
We can write this torus as 
\begin{equation}H=(H_0 \times \torus)/\Xi \subset (G_0 \times \torus)/\Xi ,
\end{equation}
where $H_0 \subset G_0$ is a maximal torus of $G_0$.  The possible charges
are described as group homomorphisms $\vartheta:H\to U(1)$, or, 
writing\footnote{Here, $\mathfrak{h}$ is the Lie algebra of $H$, which is
a finite dimensional real
vector space with trivial Lie bracket.} 
$H=\mathfrak{h}/\pi_1(H)$, as 
group homomorphims $\vartheta:\pi_1(H)\to\mathbb{Z}$
(extended from $\pi_1(H)$ to $\mathfrak{h}$ by extending scalars from $\mathbb{Z}$
to $\mathbb{R}$).

The set of possible charges 
\begin{equation}
\Lambda_w=\operatorname{Hom}(\pi_1(H),\mathbb{Z})
\end{equation}
forms the {\em weight lattice}\/ (or {\em charge lattice}\/)
inside the real vector space
$\mathfrak{h}^*=\operatorname{Hom}(\mathfrak{h},\mathbb{R})$,
and each finite-dimensional representation involves a finite number
of such {\em weights}, which are the simultaneous $H$-eigenvalues
under the representation. 
(In this context, a {\em lattice}\/ inside a real vector space is a free
$\mathbb{Z}$-module generated by a basis of the vector space.)
The dual of the weight lattice
\begin{equation}
\Lambda_{cw}=\operatorname{Hom}(\Lambda_w,\mathbb{Z})
\end{equation}
forms the {\em coweight lattice}\/ which coincides with $\pi_1(H)$.

The weights which occur as simultaneous $H$-eigenvalues
in the adjoint representation of $G^0$ are known as the {\em roots of
$G^0$}, and span a sublattice $\Lambda_r$ of $\Lambda_w$
known as the {\em root lattice.}\/ Notice that since $\torus$ lies
in the center of $G^0$, the adjoint representation of
$G^0$ maps $\torus$ to the identity, which implies that $\Lambda_r$ is a sublattice
of $(\Lambda_w)_0=\operatorname{Hom}(\pi_1(H_0),\mathbb{Z})$, 
the weight lattice of $G_0$.   In fact, $\Lambda_r$ has finite index in
$(\Lambda_w)_0$, and $(\Lambda_w)_0/\Lambda_r \cong \prod_{i=1}^k Z_i$,
where $Z_i$ is the center of the simple factor $G_i$ of $G_0$.

To specify a representation of $G^0$, we specify a representation of
each $G_i$ as well as a representation of $\torus$, and tensor them together;
this will give a representation of $G^0$ provided that $\Xi$ acts
trivially.  
The corresponding representation of $G_i$ determines a representation of its center
$Z_i$, and these -- together with the representation of $\torus$ -- determine the action
of $\Xi$.
 
The fundamental group of $G_0\times \torus$ is
\begin{equation}
 \pi_1(G_0\times \torus)=\pi_1(\torus)=\mathbb{Z}^r
\end{equation}
where $r=\dim \torus$, and its universal cover is $G_0 \times \mathbb{R}^r$,
so there is an exact sequence of abelian groups
\begin{equation}
 0 \to \pi_1(G_0 \times \torus) = \mathbb{Z}^r \to \pi_1(G^0) \to \Xi\to0
\end{equation}

Thus, a maximal free subgroup of $\pi_1(G^0)$ determines the coweights
of representations, and the finite quotient $\Xi$ determines the
compatibility conditions.  When $\Xi$ does not act on $\torus$, then it is
isomorphic to the torsion subgroup of $\pi_1 (G^0)$; on the other
hand, in the explicit examples described above where $\Xi$ acts
nontrivially on $\torus = U(1)$, $\pi_1 (G^0) =\Z$ and there is no torsion
subgroup.

\end{document}